# Life as the Only Reason for the Existence of $N_2$-$O_2$-Dominated Atmospheres

Laurenz Sproß[1,2], Manuel Scherf[1], Valery I. Shematovich[3], Dmitry Bisikalo[3], Helmut Lammer[1]

[1] Austrian Academy of Sciences, Space Research Institute, Graz, Austria
[2] Institute of Physics, University of Graz, Austria
[3] Institute of Astronomy, Russian Academy of Sciences, Moscow, Russia

Corresponding Author:
manuel.scherf@oeaw.ac.at
Austrian Academy of Sciences, Space Research Institute,
Schmiedlstr. 6, 8042 Graz, Austria

## ABSTRACT

The Earth's $N_2$-dominated atmosphere is a very special feature. Firstly, $N_2$ as main gas is unique on the terrestrial planets in the inner solar system and gives a hint for tectonic activity. Studying the origins of atmospheric nitrogen and its stability provides insights into the uniqueness of the Earth's habitat. Secondly, the coexistence of $N_2$ and $O_2$ within an atmosphere is unequaled in the entire solar system. Such a combination is strongly linked to the existence of aerobic lifeforms. The availability of nitrogen on the surface, in the ocean, and within the atmosphere can enable or prevent the habitability of a terrestrial planet, since nitrogen is vitally required by all known lifeforms. In the present work, the different origins of atmospheric nitrogen, the stability of nitrogen dominated atmospheres, and the development of early Earth's atmospheric $N_2$ are discussed. We show why $N_2$-$O_2$-atmospheres constitute a biomarker not only for any lifeforms but for aerobic lifeforms, which was the first major step that led to higher developed life on Earth.

# 1. PROPERTIES AND EVOLUTION OF $N_2$-DOMINATED ATMOSPHERES

## 1.1 Stability

An extrasolar planet with a mass and size similar to the Earth is often regarded to be habitable when it lies within the habitable zone of its host star [e.g. 1, 2]. Such an assumption often implies that the planet's atmosphere will be stable against loss to space or into its surface. Therefore, the habitability of an exoplanet is often conjectured even though the evolution of its atmosphere remains highly unclear. Moreover, climate models [3–9] often presuppose $N_2$-dominated atmospheres without considering their stability against the XUV irradiation and mass loss of its host star. To understand, however, whether an exoplanet might be habitable or not, one has to critically assess the stability of its atmospheric inventory.

Since the habitable zone environments of the distinct spectral classes such as M- and G-type stars are entirely different, principally the evolution of their incident stellar wind and short-wavelength irradiation [10, 11, 12 for M-type stars, 13, 14, 15 for G-type stars], one cannot simply apply the evolution of our home planet's atmosphere analogously to low-mass exoplanets around other stars. As studies for solar-like stars have shown [14, 15], their mass loss and XUV flux is coupled with the evolution of their stellar rotation rate. As one can read from Figure 1a, which shows the evolution of XUV fluxes of solar-like stars for a slow, moderate and fast rotator, after about 1 billion years all evolution tracks converge making it challenging to reconstruct the rotational history of a star. Figure 1b shows the evolution of the X-ray flux in the middle of the habitable zone for a moderate rotator with 0.25, 0.5, 0.75, and 1.0 $M_{Sun}$ [12]. It is clearly visible that the X-ray luminosities of lower-mass stars decrease significantly slow than solar-like stars.

By modeling the evolution of primordial hydrogen atmospheres some insights into the host stars' activity history can be gained, as recently illustrated by e.g. Kubyshkina et al. [16] for CoRoT-7 and HD 219134, and by Lammer et al. [17] for the Sun. By reproducing the fractionation of noble gas isotope ratios in the present-day atmospheres of Venus and the Earth via the escape of their primordial hydrogen-dominated atmospheres, Lammer et al. [17] suggest that the Sun must have been a slow, or at least a slow-to-moderate rotator. This is also in agreement with another study that investigated the response of the terrestrial nitrogen-dominated atmosphere to the increasing XUV irradiation during the Archean eon [18], as well as with research on the evolution of the Lunar surface composition [19].

There are only a few studies that thoroughly investigated the response of XUV irradiation onto a nitrogen-dominated atmosphere [18, 20–23 for Earth, 24 for Titan]. They found that such an atmosphere starts to adiabatically expand for a certain XUV flux range; for the Earth this is typically around 5 to 6 times the present-day solar XUV flux ($XUV_\odot$). Such a hydrdynamic expansion of the upper atmosphere is typically accompanied by a strong increase in atmospheric bulk flow, with the gas approaching escape velocities at about 10 $XUV_\odot$ [21], see Figure 2, which leads to high thermal escape rates. Johnstone et al. [25] showed that a nitrogen-dominated atmosphere of a planet with a similar mass than the Earth forms a transonic hydrodynamic Parker wind around very active stars with an outflow velocity that exceeds the escape velocity. For 60 $XUV_\odot$ an $N_2$-$O_2$ atmosphere of 1 bar would be lost within only 0.1 Myr. For a slow rotator such an XUV flux would be already reached after about 20 Myr, whereas for a moderate and fast rotator 60 $XUV_\odot$ would be approached not before around 250 Myr and 550 Myr, respectively, implying that a nitrogen-dominated atmosphere would not have been stable on early Earth.

This result also has a profound effect on the stability of such atmospheres around K- and M-stars since they show a much slower decline in their XUV irradiation. M stars stay very active likely for a billion years or even longer [12, e.g. 26]. Any nitrogen-dominated atmosphere at exoplanets orbiting such stars would not survive their high activity phase and the planet would probably lose most of its volatiles before an Earth-like atmosphere could theoretically be sustained. Johnstone et al. [18] found that even a nitrogen-dominated atmosphere with 10% $CO_2$ becomes unstable for $L_x \sim 5$ to 10 erg $s^{-1}$ $cm^{-2}$. While this value would be reached for a moderate rotator with 1 $M_{Sun}$ after about 1 billion years, it takes several billion years, or even longer, for M stars (see Figure 1b).

For smaller bodies, this effect would be even much more pronounced: Erkaev et al. [23] simulated the evolution of Titan's nitrogen-dominated atmosphere over time and found a strong increase in the exobase altitude for significantly higher XUV fluxes (see Figure 2). At the orbit of Saturn, which is about 10 AU from the Sun, Titan's atmosphere would be expanded up to more than 2 $R_{Titan}$ for 100 $XUV_{\odot}$ and up to about 4.5 $R_{Titan}$ for 400 $XUV_{\odot}$, which would be accompanied by extreme atmospheric escape of up to $4.6 \times 10^{29}$ particles per second. Even though these high fluxes only dominated the history of the early solar system (Figure 1), this has significant consequences, if Titan would be set to Earth's orbit at 1 AU: As can be seen in Figure 2, Titan's nitrogen-dominated atmosphere would experience strong atmospheric escape already at present day (100 $XUV_{\odot}$ at Titan's orbit equals approximately 1 XUV⊙ at Earth's orbit) with thermal escape rates of $1.4 \times 10^{28}$ $s^{-1}$, if one applies the results of Erkaev et al. [23]. The real escape rates, however, would likely be higher due to the elevated equilibrium temperature at 1 AU compared to 10 AU, for which the simulation runs of Erkaev et al. [23] were performed. A nitrogen-dominated atmosphere around a Titan-like body would, consequently, not be stable in the solar habitable stone, not even for very low XUV fluxes. From this perspective, Titan can hardly be compared with the Earth.

Besides thermal escape, nitrogen-dominated atmospheres are also susceptible to strong non-thermal loss processes if the XUV flux from the Sun, or more generally from its host star, is significantly elevated. By using the atmospheric profiles of Tian et al. [21], Lichtenegger et al. [27] have shown that for 20 XUV⊙ an atmosphere with the present-day composition would have been completely eroded within just a few million years by ion-pickup through the then much stronger ancient solar wind [13]. In addition, Kislyakova et al. [28] found very strong polar escape rates of $N^+$ ($5.1 \times 10^{27}$ $s^{-1}$) and O+ ($4.3 \times 10^{27}$ $s^{-1}$) even for much lower fluxes of only about 5 XUV⊙. Such high polar loss rates of ions are also supported by another study [29], which found that exoplanets around K and M stars can lose their entire nitrogen atmosphere within a few 10s to 100s of Myr through ion escape that is induced by the strong solar wind and XUV irradiation in the habitable zones of such stars. In addition, CMEs may further erode exoplanetary atmospheres around M stars very efficiently [30–33]. Strong thermal escape in addition to such pronounced non-thermal escape strongly indicates that a nitrogen-dominated atmosphere would not have been stable even for putatively rather ordinary XUV fluxes in the range of 10 to 20 $XUV_{\odot}$ or even lower.

These findings are in agreement with a very recent study by Johnstone et al. [18], which investigated the effect of different $N_2$-$CO_2$ mixing ratios onto the stability of nitrogen atmospheres. As already found by other research [20, 27, 34, 35], $CO_2$ acts as an infrared cooler in the upper atmosphere leading to a decrease in exobase level. Consequently, Johnstone et al. [18] studied the effect of different XUV fluxes onto atmospheres with various $N_2$-$CO_2$ mixing ratios, as can be seen in Figure 3 for an evolving slow rotating Sun. Their results strongly indicate that even during the Archean eon much higher values of $CO_2$ must have been in the atmosphere to keep its nitrogen stable. In addition, the Sun also should have been born as a slow or (rather slow) moderate rotator to prevent the terrestrial atmosphere from escape during the Archean [18].

To summarize, the present-day terrestrial $N_2$-$O_2$-atmosphere would not have been stable in the Hadean and early Archean and must, therefore, have grown to its present mass later-on. Moreover, K and in particular M stars are likely to be no hospitable place for the evolution of $N_2$-$O_2$-atmospheres. They can only be stable and evolve within the respective habitable zones after several billion years depending on stellar mass and rotation rate; stars below 0.4 $M_{Sun}$ likely never cross a threshold of $L_x < 10$ erg $s^{-1}$ $cm^{-2}$ [12]. It might be questionable that an exoplanet around such stars can keep their volatiles for such a long timeframe. But to answer this question further studies are needed.

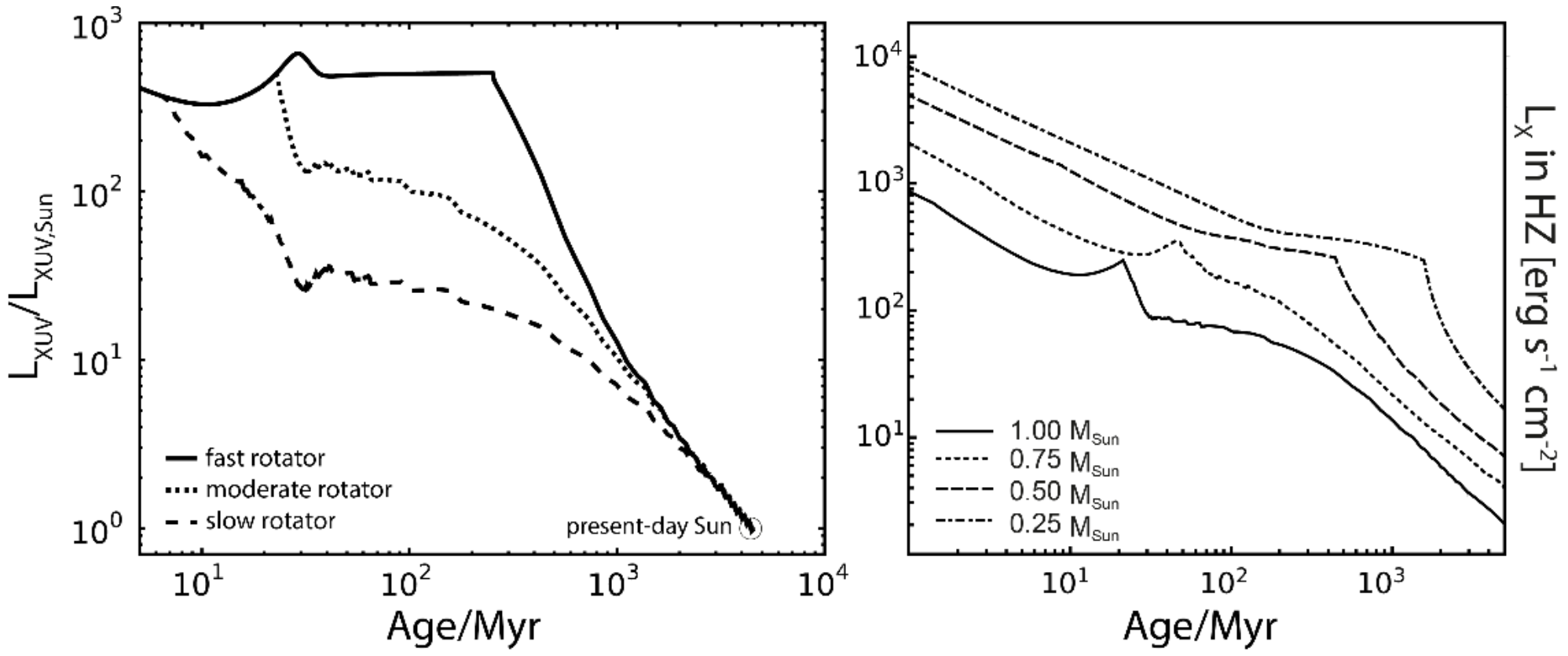

Figure 1: a) XUV flux evolution of solar-like stars according to [15]. While slow, moderate and fast rotators show significant differences in their XUV flux early on by up to an order of magnitude, after about 1 billion years all different tracks converge towards one [figure after 15]. b) X-ray luminosity for a moderate rotator of different masses scaled to the respective habitable zones. Lower-mass stars stay active significantly longer [figure after 12].

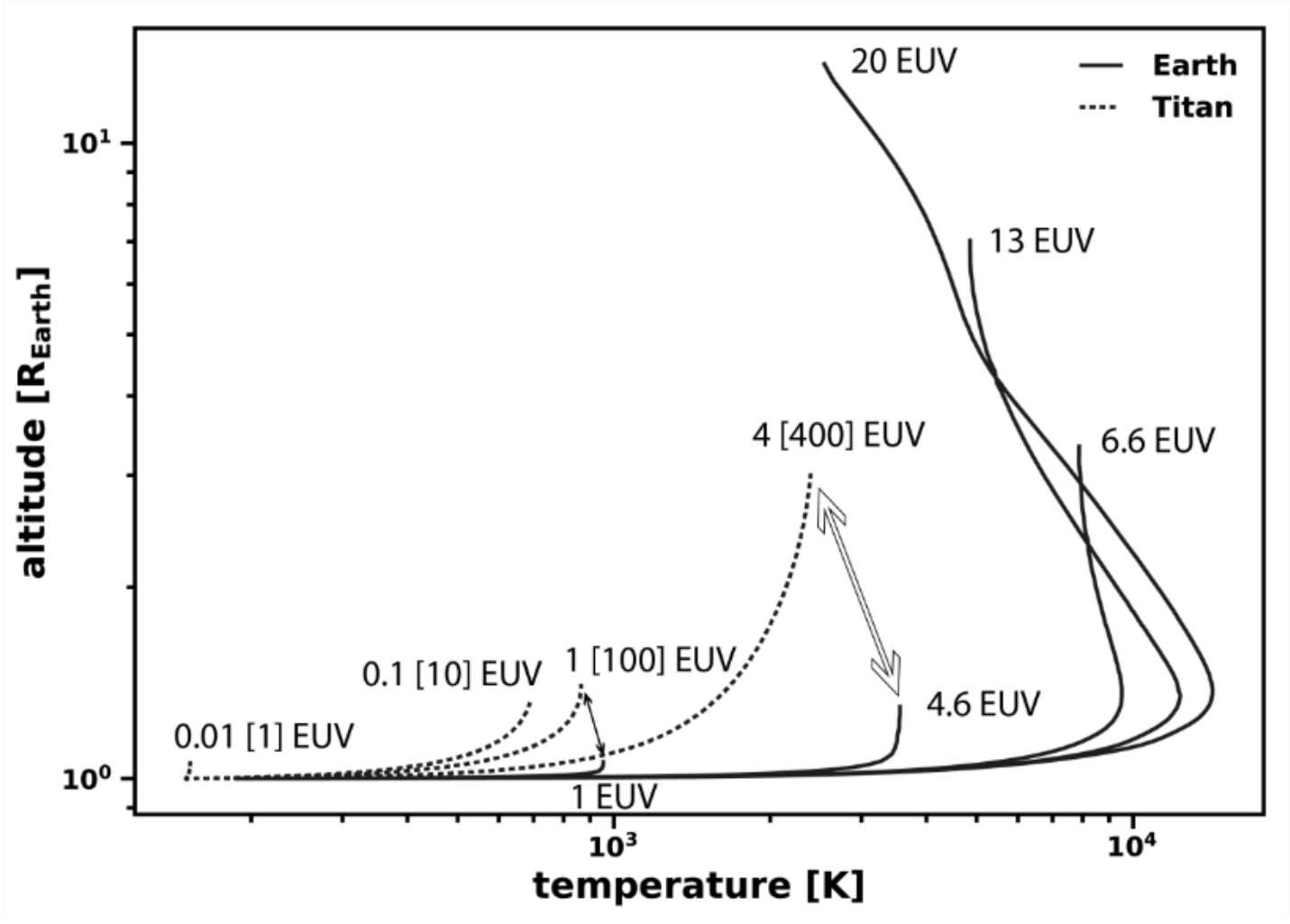

Figure 2: The response of the Earth's (present-day $CO_2$ abundance; after [21]) and of Titan's nitrogen-dominated atmosphere [after 23] against different incident XUV fluxes (exobase level on the top of the lines). Both bodies respond with significant expansion to increased XUV, but due to its significantly lower mass ($M_{Titan}/M_{Earth} \approx 0.02$), however, Titan's atmosphere likely would not even be able to survive the present-day irradiation at Earth's orbit because of strong atmospheric escape. Here, the numbers in brackets depict the XUV flux at Titan's orbit while the others indicate the flux at Earth's orbit. It has to be noted that the Titan profiles were simulated for Titan's orbit (≈10 AU), i.e. for a significantly lower equilibrium temperature of $T_{eq}$ = 150 K (for Earth's orbit $T_{eq} \approx$ 250 K). Therefore, Titan's atmosphere would be even more susceptible to atmospheric escape if the satellite would suddenly be transferred to 1 AU.

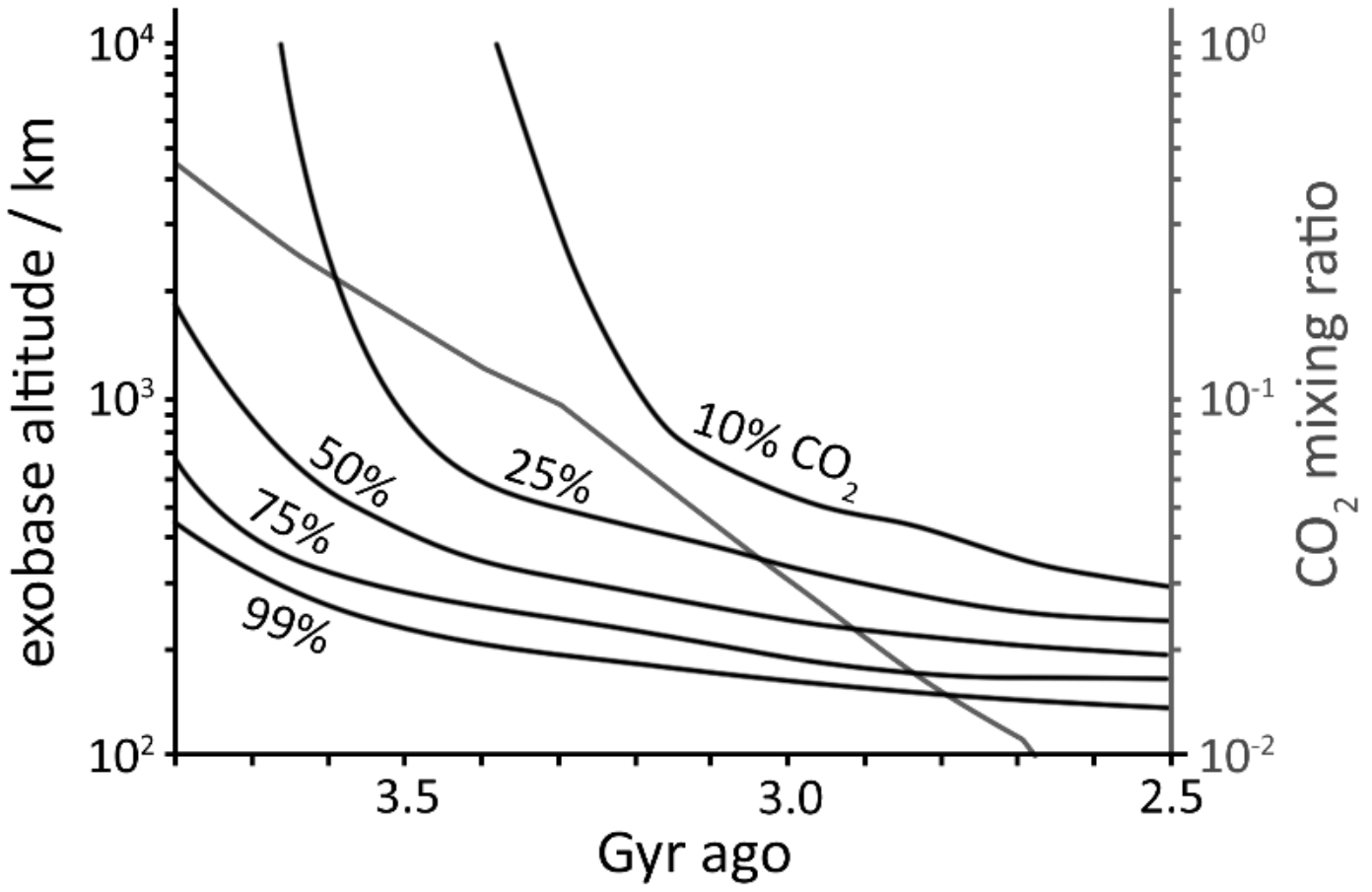

Figure 3: Exobase altitudes over time for the XUV flux evolution of a slow rotating Sun and $CO_2$ mixing ratios given in % are plotted in black and relate to the left axis. The minimum $CO_2$ mixing ratio for the atmosphere to be stable for a slow rotating Sun is drawn in grey and relates to the right axis (Figure adopted from [18]).

## 1.2 Isotopic nitrogen ratios and their origin

The isotopic ratio of $^{14}N/^{15}N$ in the atmospheres of Earth, Titan, but also Triton, and Pluto, can tell us something about their building blocks, since different reservoirs in the solar system can be divided by their various fractionations [36]. However, one has to be aware that isotopic ratios can change over time due to a diverse set of different factors such as atmospheric escape [23, e.g. 37, 38], photochemistry [e.g. 39, 40], chemical, geological and biological processes [41, e.g. 42–45]. Therefore, it is not straightforward to derive the origin of atmospheric nitrogen from its present-day fractionation since the aforementioned processes have to be taken into account.

The following different nitrogen reservoirs can currently be distinguished in the solar system through their isotopic fractionation:

1. The solar wind and Jupiter: The solar wind ($^{14}N/^{15}N = 440.5 \pm 5.8$ [46]) and Jupiter's atmosphere (435 ± 65 [47]) show the lightest isotopic nitrogen ratios in the solar system. They are representative for the solar value and primordial $N_2$ within the protoplanetary nebula (PSN; e.g. [48]).

2. Cometary ices: $NH_3$ ices in comets constitute the heaviest nitrogen reservoir currently known in the solar system. Spectral observations of the Jupiter family and Oort cloud members retrieved values of $^{14}N/^{15}N = 127 \pm 32$ [49] which is in good agreement with another study that analyzed optical spectra of 18 additional comets from various dynamical groups in the outer solar system ($^{14}N/^{15}N = 135.7 \pm 5.9$ [50]). This reservoir might be the main source of Titan's atmospheric nitrogen [23, e.g. 37].

3. Refractory organics: Interplanetary dust particles and carbonaceous chondrites from the outer solar system can contain complex refractory organics which include nitrogen with a wide variety of isotopic ratios and a mean value of $^{14}N/^{15}N \approx 231$ [e.g. 51, 52]. Refractory organics might have also played a role in the formation of Titan's atmosphere [52].

Chondrites, a diverse group of primitive unprocessed meteorites, form an additional reservoir, but with its nitrogen being a mixture of reservoirs 2 and 3. This reservoir can be divided into different subgroups such as carbonaceous, ordinary and enstatite chondrites. They have a bulk ratio of $^{14}N/^{15}N = 259 \pm 15$ [53]. The terrestrial atmospheric nitrogen likely originates from these building blocks and also the Martian interior and the atmosphere of Venus show comparable values.

For the Earth, the present-day atmospheric isotopic nitrogen ratio is $^{14}N/^{15}N = 272$ [e.g. 48] which is very close to the chondritic bulk ratio. The reverse of this value also serves as the nitrogen isotope standard $(^{15}N/^{14}N)_{standard} = 3.676 \times 10^{-3}$ for the so-called stable isotope delta notation

$$\delta^{15}\mathrm{N} = \left(\frac{(^{15}\mathrm{N}/^{14}\mathrm{N})_{\mathrm{sample}}}{(^{15}\mathrm{N}/^{14}\mathrm{N})_{\mathrm{standard}}} - 1\right) \cdot 1000$$

with $\delta^{15}N$ as deviation of the sample from the standard in ‰. Interestingly, the Earth shows a slight isotope disequilibrium between internal and surface reservoirs [e.g. 54], with atmospheric nitrogen ($\delta^{15}N = 0$ ‰) being slightly enriched in $^{15}N$ compared to the upper ($\delta^{15}N < -5$ ‰) and deep mantle ($\delta^{15}N < -40$ ‰). The latter is believed to be a relic from the primordial mantle representing nitrogen isotope fractionation ≈4.56 Gyr ago. The highest $^{15}N$-enrichment, however, can be found in the terrestrial crust which, at present-day, has a value of about $\delta^{15}N \approx +6$ ‰. Sedimentary analysis of ancient crust before 2.7 Gyr ago revealed an enrichment of up to $\delta^{15}N \approx +13$ ‰ [42]. Lammer et al. [55] argued that outgassing of $N_2$ enriched in $^{14}N$ from the deep mantle might point to an even lower $^{14}N/^{15}N$ ratio in the ancient atmosphere before the Great Oxidation Event.

The reason for the nitrogen isotope disequilibrium can be manifold and could have been caused by enriching $^{15}N$ through atmospheric escape, biological processes, mantle-core segregation, late delivery of $^{15}N$-enriched volatiles that formed the atmosphere or by devolatilization and recycling of sedimentary rocks [e.g. 54]. Biological processes [56], on the other hand, can either preferentially release the lighter isotope into the atmosphere through denitrification and anammox ($\delta^{15}N \approx$ -10 to +25 ‰) or the heavier through fixation ($\delta^{15}N \approx$ 0 to -4 ‰).

Atmospheric escape can be another reason for the enrichment of $^{15}N$ in the terrestrial atmosphere since it preferentially removes the lighter isotope [see e.g. , 38 for an extensive discussion]. This process is responsible for a significant fractionation of $^{14}N/^{15}N$ at Mars [57–60], as can be seen in the remarkable isotopic disequilibrium between the atmosphere ($\delta^{15}N$ = +620 ± 160 ‰ [61, 62]) and its mantle ($\delta^{15}N$ = -30 ‰ [63]). This indicates that up to 90 % of the Martian atmospheric nitrogen could have been lost over time through different loss processes such as sputtering or photochemical escape [64].

On Earth, however, the small nitrogen isotope disequilibrium indicates that atmospheric escape might have only played a minor role. The fractionation of an isotope ratio can be related with the total amount lost from an atmosphere through the so-called Rayleigh distillation relationship through

$$\frac{n_0}{n} = \left(\frac{R}{R_0}\right)^{\frac{1}{1-f}},$$

where $n_0$ and $n$ are the initial and the current inventory of an atmospheric gas, $R_0$ and $R$ are the initial and the current isotopic fractionation, and $f$ is defined as fractionation factor; for $f > 1$ the heavier and for $f < 1$ the lighter isotope is preferentially removed. Even though losses to space always yield $f \leq 1$, one has to distinguish between thermal and non-thermal escape, because these processes show different fractionation efficiencies. Thermal escape can be further divided into hydrodynamic and Jeans escape, of which the latter yields low escape rates but significant fractionation and can be defined as

$$f = \sqrt{\frac{m_1}{m_2}}\left[e^{(\lambda_1-\lambda_2)}\frac{(1+\lambda_2)}{(1+\lambda_1)}\right],$$

with $m_1$ and $m_2$ as the lighter and heavier isotopes, and $\lambda_{1,2} = GMm_{1,2} / (k_B T_{exo} r_{exo})$ as Jeans escape parameter. Here, G is the gravitational constant, $k_B$ the Boltzmann constant, $M$ the mass of the planet or satellite, $T_{exo}$ the temperature at the exobase, and $r_{exo}$ is the radial distance to the exobase. A Jeans escape parameter of $\lambda \geq 6$ indicates that the escape flux is purely dominated by Jeans escape; whereas it is purely hydrodynamic for $\lambda \leq 2$–3. In the latter hydrodynamic case escape can remove a substantial amount of volatiles but with almost no fractionation since for very low values of λ the fractionation factor converges towards

$$f \approx \sqrt{m_1/m_2}.$$

Finally, non-thermal escape fractionates according to

$$f = \exp\left[\frac{g(r_{\text{exo}} - r_{\text{h}})(m_2 - m_1)}{k_{\text{B}} T_{\text{exo}}}\right].$$

Here, $g$ is the gravitational acceleration and $r_h$ is the radial distance of the homopause. It has to be noted, however, that for strongly extended atmospheres, $g$ will significantly change between $r_h$ and $r_{exo}$ and an average value of $g$ might then be more realistic.

If one assumes that the isotope disequilibrium at Earth of $\delta^{15}N$ = 40 ‰ is entirely due to atmospheric escape, then one can simply estimate the total amount lost from the atmospheric inventory through thermal and non-thermal escape with the equations above. However, there are several uncertainties that are currently not known, such as the evolution of the terrestrial atmospheric $pN_2$ and its exobase altitude and temperature which are both relevant for the Jeans escape parameter $\lambda$.

Figure 4 shows the amount of nitrogen that should have been lost from the terrestrial atmosphere, if Earth's isotope disequilibrium of $\delta^{15}N \approx 40$ ‰ were purely ascribed to atmospheric escape and in dependence of different atmospheric $pN_2$ at the time of fractionation. Let's arbitrarily consider that at 2.7 Gyr ago the partial pressure of nitrogen was about ≈200 mbar and the fractionation of $^{14}N/^{15}N$ stopped around this time due to atmospheric escape becoming negligible. In such a case, ≈600 mbar nitrogen might have been outgassed later-on from the interior to build up the present-day atmosphere, meaning that the initial atmospheric fractionation must have been higher, i.e. $\delta^{15}N \approx +120‰$, if the subsequently outgassed $N_2$ was unfractionated. Here, Figure 4a shows non-thermal and thermal (i.e. Jeans) escape for the cases of 4.6 $XUV_{\odot}$ and 6.6 $XUV_{\odot}$ (see also Figure 2, and [21]. The latter case is clearly in the Jeans escape regime, while 4.6 $XUV_{\odot}$ can already be considered to be in a transition phase between Jeans and hydrodynamic escape due to $\lambda \approx 4$. Figure 4b, on the other hand, shows purely hydrodynamic escape (such as the cases 13 $XUV_{\odot}$ and 20 $XUV_{\odot}$ from Figure 2). This illustrates that Earth could have lost a huge amount of $N_2$, if it was lost hydrodynamically early-on. Also for very low initial $pN_2$, the loss could have been substantial even for stable atmospheres. In any other case, only about 100 mbar through non-thermal or 50 mbar to 200 mbar through thermal escape might have been lost into space after an $N_2$-dominated atmosphere could have been stable at Earth since the early to mid-Archean eon [see 18]. But as illustrated in Section 1.1, a nitrogen-dominated atmosphere would not have been stable against high XUV fluxes, meaning that any potential initial nitrogen in the atmosphere must have been lost completely and built-up again later through outgassing.

Besides Earth and Mars, loss to space also altered the isotopic ratios in the atmosphere of Titan, although less significant than at Mars [23, 37]. However, photolysis of $N_2$ and subsequent sequestration into HCN, a fractionation process that preferentially removes the heavier isotope, counteracted against the enrichment of $^{15}N$ in Titan's atmosphere [e.g. 39, 40]. Even though loss to space might have been able to remove a substantial amount of Titan's nitrogen reservoir, the isotopic ratio did not change substantially due to these counteracting fractionation processes. At present-day, Saturn's largest satellite shows a strong enrichment of $^{15}$N compared to Earth with $\delta^{15}$N ≈ +622 ‰ [65], while the initial value was recently estimated to be $\delta^{15}$N ≥ +600 ‰ [23]. Although earlier studies hypothesized that Titan's atmosphere was composed out of the same building-blocks than the terrestrial [e.g. 66], these new findings together with other recent studies [37, 39, 52, 67] strongly highlight that both atmospheres mainly formed from different reservoirs. Titan's nitrogen is likely to have originated from ammonia ices and refractory organics [23, 52], while the Earth's was mainly delivered through primitive meteorites such as carbonaceous chondrites [68].

Besides the origin from different sources, nitrogen can also be considered to be available in higher abundances in the outer solar system than in habitable zones. Comets as well as carbonaceous chondrites (CC), the main carrier of nitrogen-bearing refractory organics and ammonia ices, have their origin in the outer solar system beyond the ice line. Even though some amount of these volatile rich bodies were scattered into the inner solar system – Earth accreted about 2 – 5 % CCs and comets [17, 68] – their main feeding zone lies beyond the habitable zone. The temperature in the HZ is, in addition, too high to enable condensation and direct accretion of $N_2$ onto a planetary embryo. It can, therefore, be expected that exoplanets and exomoons outside the HZ and farther away from their host star will, like Titan, generally have a larger initial nitrogen reservoir compared to planets within the HZ. This can be seen as the second important factor why the Earth, or more general, an Earth-like habitat – a planet that is within the HZ and surrounded by a nitrogen-dominated atmosphere – cannot be compared with an icy body such as Titan.

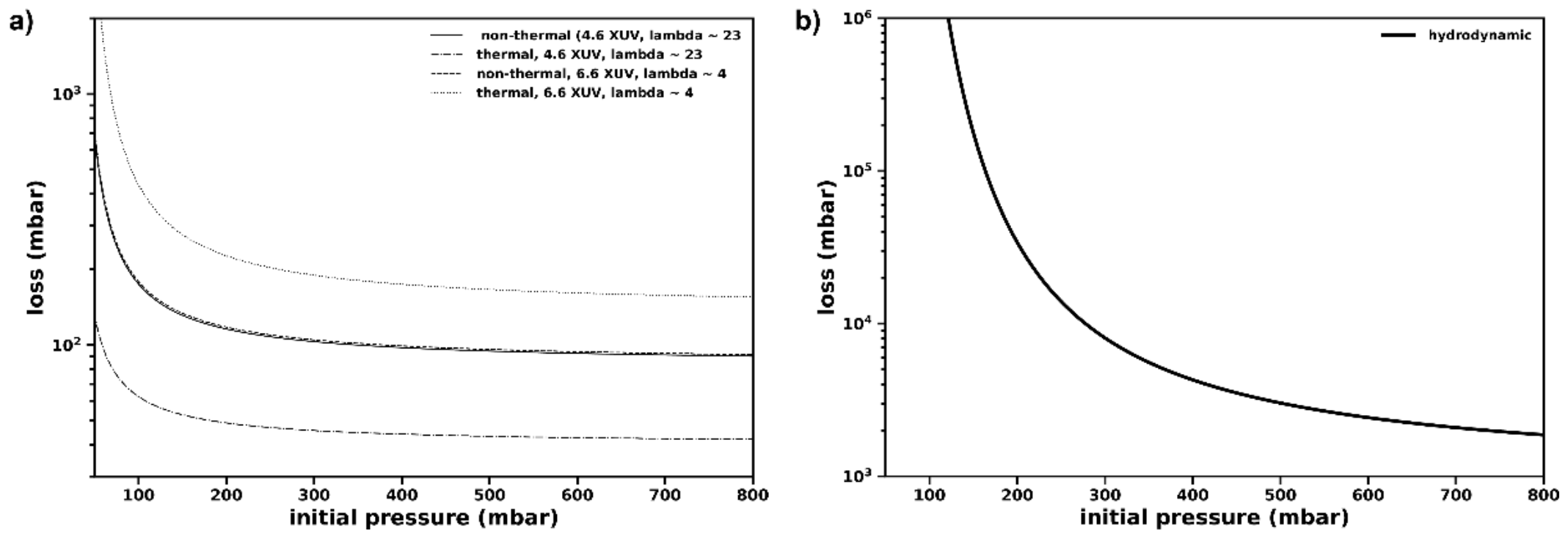

Figure 4: The amount of nitrogen that has to be lost through Jeans and non-thermal escape (a) and hydrodynamic escape (b) for different initial $pN_2$ to account for the terrestrial isotope disequilibrium of $\delta^{15}N \approx 40‰$. While (a) shows 4.6 $XUV_\odot$ and 6.6 $XUV_\odot$ (see also Figure 2), (b) illustrates any hydrodynamic case such as 13 $XUV_\odot$ of [21].

### 1.3 Buildup and depletion of atmospheric $N_2$

Since dinitrogen is bond very strongly through a triple-bond N≡N and is also very volatile, it is not easy to transform it into reactive compounds and eventually deplete it from an atmosphere through deposition. Sufficient energy for the fixation of $N_2$ can be provided for instance by impactors, by UV light, by cosmic rays, or within lightning strokes [69, e.g. 70, 71]. On Earth, nitrogen fixation is mainly performed by cyanobacteria, which are one of the oldest microbial species and have transformed $N_2$ into $NH_3$ at least since 3.5 Gyr, maybe since more than 3.7 Gyr ago [e.g. 72, 73].

While on present Earth nitrogen is efficiently fixed as well as released by lifeforms – 1 or 2 orders more efficiently than abiotic processes, see e.g. Fowler et al. [74] or Lammer et al. [69, their appendix] – under the reducing atmospheric conditions in the Archean, a different form of the nitrogen cycle was in place, wherein an efficient biogenic release of nitrogen was not possible (see Figure 5). Such a recycling of previously fixed nitrogen through denitrification and/or anammox is only possible on the basis of oxidized nitrogen compounds and therefore requires an active aerobic part of the nitrogen cycle. Thus, prior to the oxidation of the Earth's atmosphere and ocean, nitrogen must have been either volcanically outgassed very strongly or efficiently chemically recycled in order to balance the atmospheric abundance, or alternatively, underwent degradation, since the present balancing-mechanism is not older than the GOE. A detailed discussion of the geobiological nitrogen cycle in relation to nitrogen atmosphere buildup is provided by Lammer et al. [69, Sec. 2]. The onset of the aerobic nitrogen cycle during the GOE transition was described by Zerkle et al. [56].

Another important factor is the gradual oxidation of the interiors, which is responsible for volcanic nitrogen outgassing at least on terrestrial planets. In silicates under reducing conditions, nitrogen is predominantly present in the form of ammonia [75]. While ammonia is fairly soluble, $N_2$ is incompatible in such minerals [45, e.g. 76]. An efficient nitrogen outgassing therefore requires an oxidation of the lithosphere, which on Earth has been ongoing continuously over the planet's lifetime due to tectonic activity. Moreover, there are strong indications that a relatively high mantle oxidation state was reached already in the time of the Moon-forming impact [77], which was not only responsible for outgassing of more $CO_2$/$H_2O$ and less $CH_4$/$H_2$ [77, 78], but also might have helped to reach a lithospheric oxidation appropriate for $N_2$ outgassing more easily at a later time. While $H_2O$ can rain out later when surface temperatures drop, $CO_2$ can be dismantled through the carbon-silicate cycle on a tectonically active body in the long run. On Venus, for example, these conditions were not fully met and therefore, there is atmospheric $N_2$ present, but $CO_2$ in much larger amounts. This complex interplay led to the assumption by Lammer et al. [69] to regard $N_2$ domination (and $CO_2$ absence) in an atmosphere as indicator for tectonic activity, which they call a "geosignature".

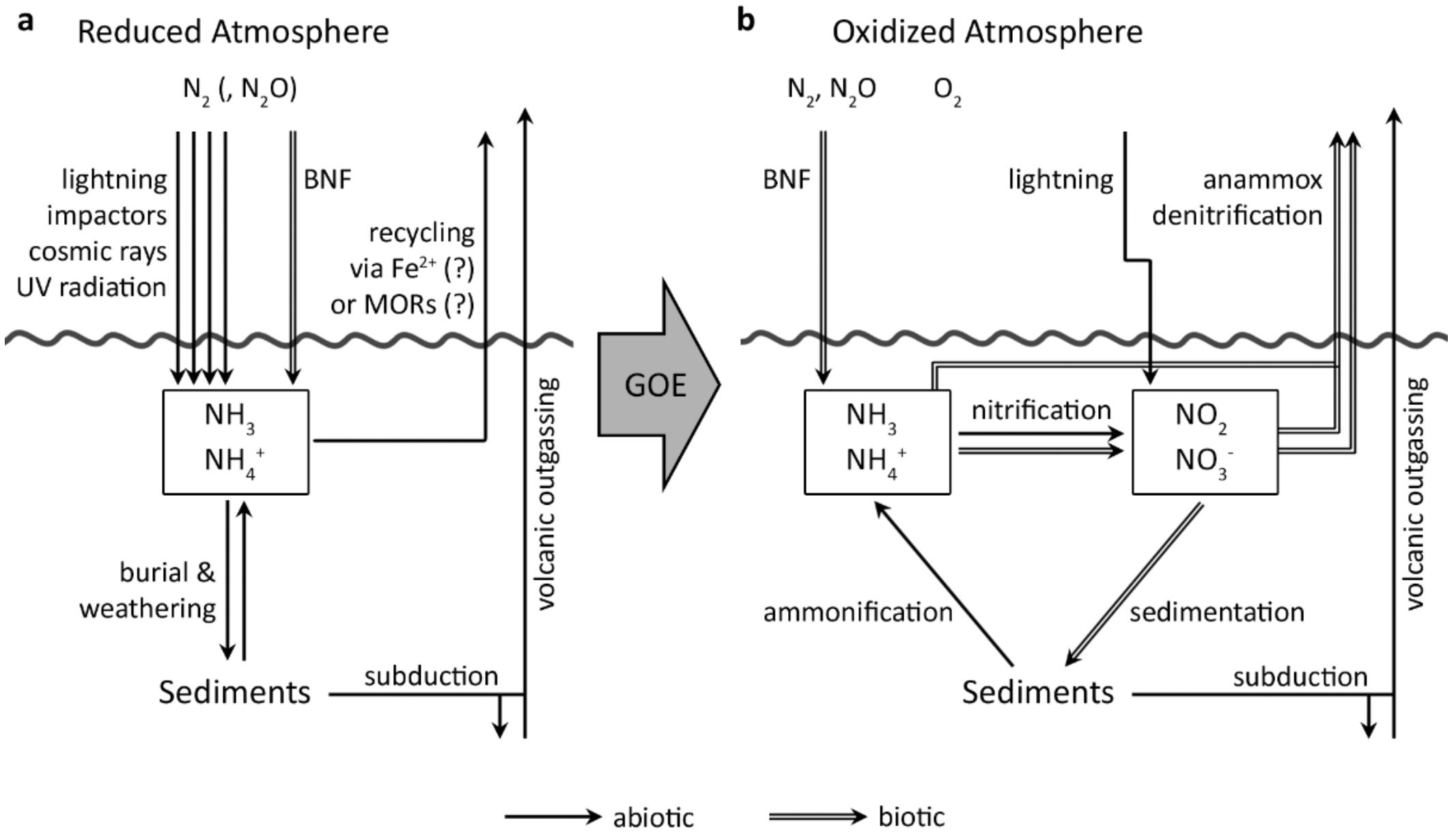

Figure 5: A simplified representation of the Earth's nitrogen cycle before and after the great oxidation event (GOE). This scheme is basically applicable to other rocky planets or moons, too. a) In the Hadean and Archean, the Earth had reducing atmospheric and marine conditions. Diverse fixation processes, abiotic and biotic nitrogen fixation (the latter is abbreviated as BNF), can lead to efficient nitrogen fixation in the form of $NH_3$, $NH_4^+$, and HCN. The possibility of a process that closes the circle by recycling fixed nitrogen back into the atmosphere is under debate (see Section 2.2), but – if existing – most likely not capable of balancing atmospheric nitrogen pressures on a planet/moon with cyanobacterial activity. After interacting with sediments, nitrogen can be partially outgassed via subduction and metamorphism, or directly released from the mantle. b) Under oxidizing conditions, a second, aerobic part of the nitrogen cycle starts to work. Now, reduced nitrogen can be transformed to $NO_2$ or $NO_3^-$ Via nitrification and thereafter released as $N_2$

into the atmosphere via denitrification or anammox. Nitrification, denitrification, and anammox are all exothermal processes, which makes them an energy source for microbes.

# 2. EVOLUTION OF EARTH'S NITROGEN ATMOSPHERE

## 2.1 State of the Art

Presently, one does not know much about nitrogen in the Archean. Recent studies by Marty et al. [79] and Avice et al. [80] show an upper limit of atmospheric $N_2$ of 1100 mbar, and 800 mbar, respectively, in the mid Archean (see Figure 6). Further, there are studies that give a total atmospheric pressure of 230 ± 230 mbar [81] and 300 mbar [82] for the late Archean. The initial atmospheric $N_2$ amount, which is in this context the remnant from the magma ocean phase in the beginnings of the planet's life, is not known. A widely accepted assumption is that nitrogen was catastrophically outgassed during this period reaching a few hundreds of mbar [83, 84], but there are indications that atmospheric nitrogen might have experienced efficient photochemistry [e.g. 85, 86] and was dissolved in the hot surface material [87–90]. The findings from Section 1.1 show that, because an atmosphere rich in $N_2$ was not stable under the Hadean solar irradiation, the absence of a strong fingerprint of atmospheric escape in the $^{14}N/^{15}N$ isotopic ratios gives evidence of a very low atmospheric $N_2$ amount during the Hadean and early Archean. Already during this period, strong abiotic fixation processes should have further minimized atmospheric $N_2$. In summary, a stable $N_2$-dominated atmosphere throughout this period was not existent, which is a strong indication for low initial atmospheric nitrogen [69]. This is also underlined in a recent study by Lehmer et al. [91] who found a minimum $CO_2$ mixing ratio of at least 70% at 2.7 Gyr ago from the fraction of oxidized iron on micrometeorite surfaces and the formation of magnetite. Combined with the above mentioned total pressure of 230 ± 230 mbar [81] at the same time, this indicates a maximum $N_2$ partial pressure of 69 ± 69 mbar.

From an aeronomical perspective, the history of atmospheric nitrogen is directly linked to that of $CO_2$, as described in Section 1.1. The minimum mixing ratio of $CO_2$ gives further indication for the maximum of atmospheric $N_2$ also in the middle and early Archean, as shown in Figure 6. On Earth, $CO_2$ was catastrophically outgassed into the atmosphere during the magma ocean's crystallization phase reaching some tens of bar [83, 92]. As previously addressed in Section 1.3, $CO_2$ was then weathered into the surface, a process that is directly linked with the Earth's special feature of long-term active tectonic activity [93, e.g. 94, 95].

All in all, both geological and aeronomical studies show a lower $N_2$ partial pressure than that of today in the Archean, which leads to the conclusion that nitrogen levels must have increased strongly in the very late Archean, during the GOE transition, or later.

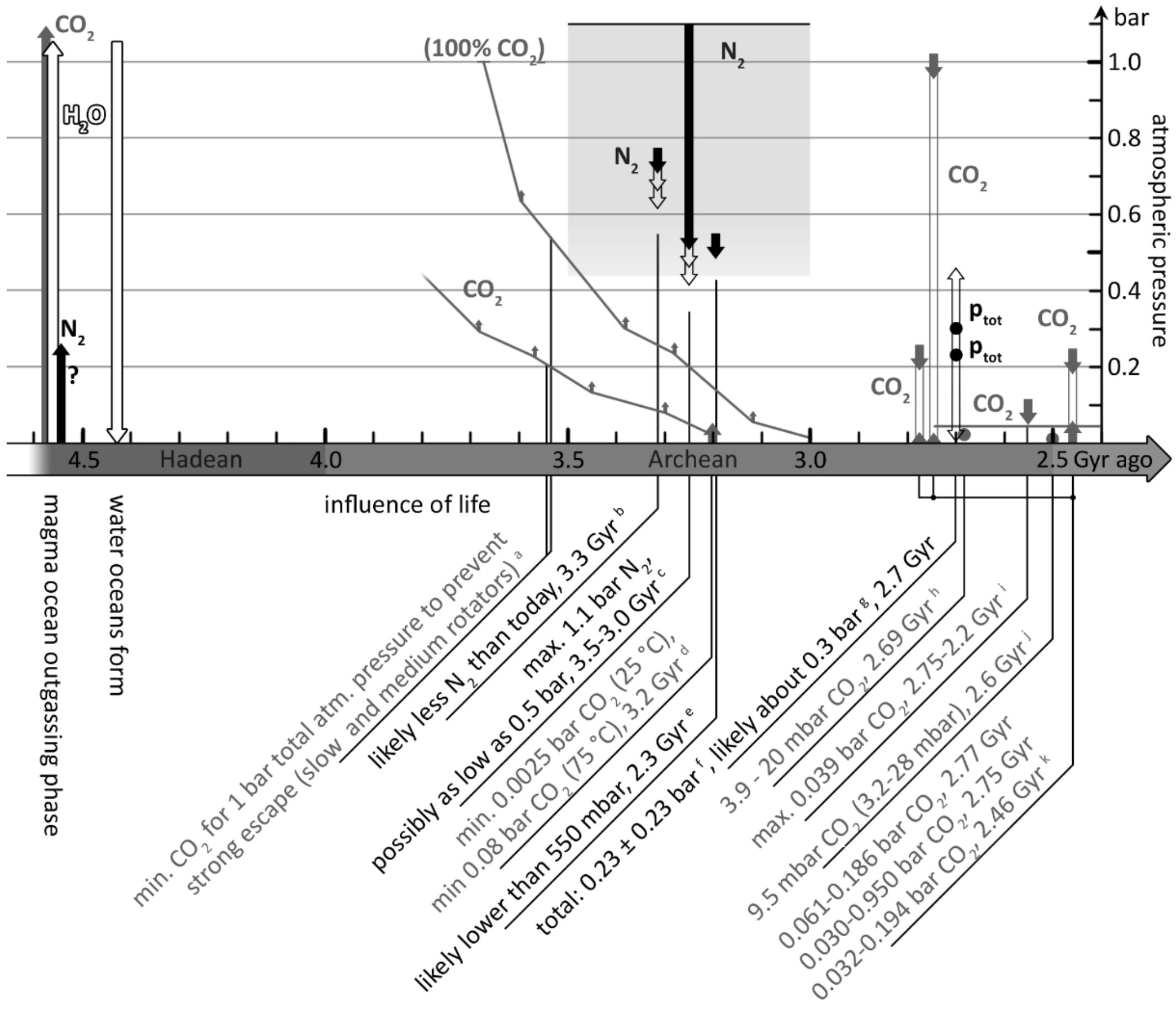

Figure 6: Data of the Earth's total atmospheric pressure $p_{tot}$ and of the $N_2$ (both black) and $CO_2$ (grey) partial pressures from sedimentary rocks, palaeosols, raindrop imprints, and calculated by a solar radiation interaction model. References: a [18], b [80], c [79], d [96], e [97], f [81], g [82], h [98], i [99], j [100], k [101]. [figure amended from 102]

## 2.2 Model approach for $N_2$ throughout the Hadean and Archean

Already 30 years ago, the rates of abiotic nitrogen fixation on early Earth were estimated to be rather high, at least as high as today [71, e.g. 103]; since then, there haven't been new insights that led to significant changes in the order of these rates. Today, the most important abiotic process consists lightning fixation, which fixes around 4 ± 1 Tg N/yr, while biological nitrogen fixation (BNF) on both land and ocean is about 198 ± 40 Tg N/yr and therefore two orders larger [69, their Table A1]. On early Earth, the influence of cyanobacteria additional to the efficient abiotic fixation could have led to $N_2$ depletion rather than buildup. To address this issue and show first-estimate $N_2$ partial pressure evolution scenarios, we developed a model for the early Earth's atmospheric nitrogen development (NDEV, see Appendix).

Model runs for an Earth on which life has never originated, are shown in Figure 7. Indeed, these runs show a continual increase of atmospheric $N_2$ throughout the Archean, but also a period of strong $N_2$ depletion in the Hadean. Further, partial pressures of more than $1 \cdot 10^9$ Tg N (≈ 200 mbar) are only reached in the very late Archean; only if one starts with a lot of remnant atmospheric $N_2$ from the magma ocean phase and rather high $CO_2$ levels, one gets $N_2$ amounts of $2 \cdot 10^9$ Tg N (≈ 400 mbar). But, as theoretically described in Section 1.1, the higher the $N_2$ levels, the faster one approaches instable regimes, in which $N_2$ escapes to space.

In Figure 8, the influence of life on the $N_2$ partial pressure is shown. BNF is so intensive that it completely depletes atmospheric nitrogen within very short periods of time, if an effective return flux is missing (Figure 8a). Such a return flux was early proposed by Kasting et al. [104] to be a nitrogen reduction in hydrothermal vent systems at mid ocean ridges. It is not clear if such a process could contribute significantly to the geochemical nitrogen cycle. Meanwhile, a more promising process was revealed. After Knauth [105] found indications for a high iron concentration in the early oceans, a recent study by Ranjan et al. [106] investigated $Fe^{2+}$-induced nitrogen outgassing (FINO), a process in which nitrogen oxides in marine environments are efficiently transformed into $N_2$ or $N_2O$ and, in most cases, ferric oxyhydroxide. The existence of this process is still under investigation, since it needs oxidized nitrogen compounds in the early oceans, which was not as common as today; similar problems were already discussed in the context of the existence of anammox [107, 108]. Nevertheless, FINO is implemented in the model runs in Figure 8b and under these conditions, stable atmospheric $N_2$ pressures can be received until the GOE. During the GOE transition, denitrification starts to contribute and to release a lot of the previously fixed and sedimented nitrogen back into the atmosphere, which leads to a boost of atmospheric $N_2$ levels, which can be seen in Figure 8. This behavior was already found by Stüeken et al. [109], who started with way more initial atmospheric $N_2$ (and unfortunately not with very low $N_2$), but still gained $N_2$ during the GOE transition (compare Figure 9). In the context of these findings, one has to mention that the NDEV model shows qualitative and not so much quantitative results; the magnitude of alterations through the model's main driver, its underlying outgassing submodel, can be seen in Figure 10).

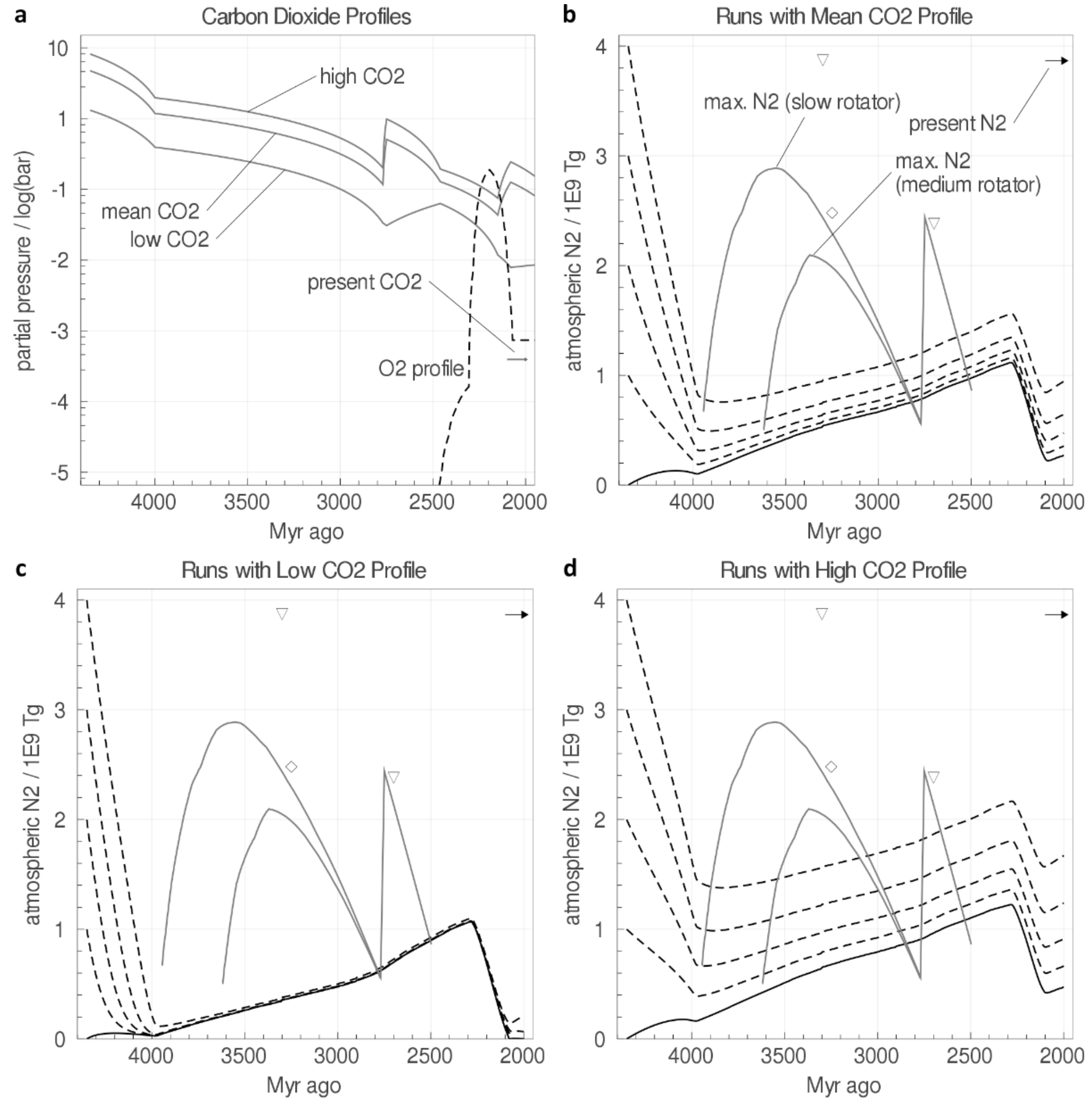

Figure 7: Model runs of the atmospheric $N_2$ partial pressure in an abiotic Earth scenario. a: Plot of the three here used $CO_2$ profiles that are following the studies from Figure 6 (see appendix for discussion). Also shown is the $O_2$ profile by Lyons et al. [110]. b, c, d: model runs for different underlying $CO_2$ profiles from panel a, and for different initial $N_2$ (here at 4350 Myr ago, see appendix). Runs with zero initial $N_2$ are drawn solid, while those with 1, 2, 3, and $4 \cdot 10^9$ Tg N are dashed. The intense fixation throughout the Hadean lowers

potential remnant $N_2$ from the magma ocean phase and makes very high partial pressures in the early Archean impossible. The maximum $N_2$ derived from the results of the Kompot-code by Johnstone et al. [20, 25] is shown in grey for a slow and a medium rotator case (see Section 1.1). The tooth-shape of these curves in the late Archean is due to the varying $CO_2$ amounts from the study by Kanzaki and Murakami [101]; here one should keep the high uncertainties of the measurements in mind. It is important to note that any kind of nitrogen escape is not considered as model process (see discussion in Section 1.2). The diamonds ◇ and triangles ▽ indicate expectation values and upper values from geological studies for atmospheric $N_2$ pressures (see Figure 6).

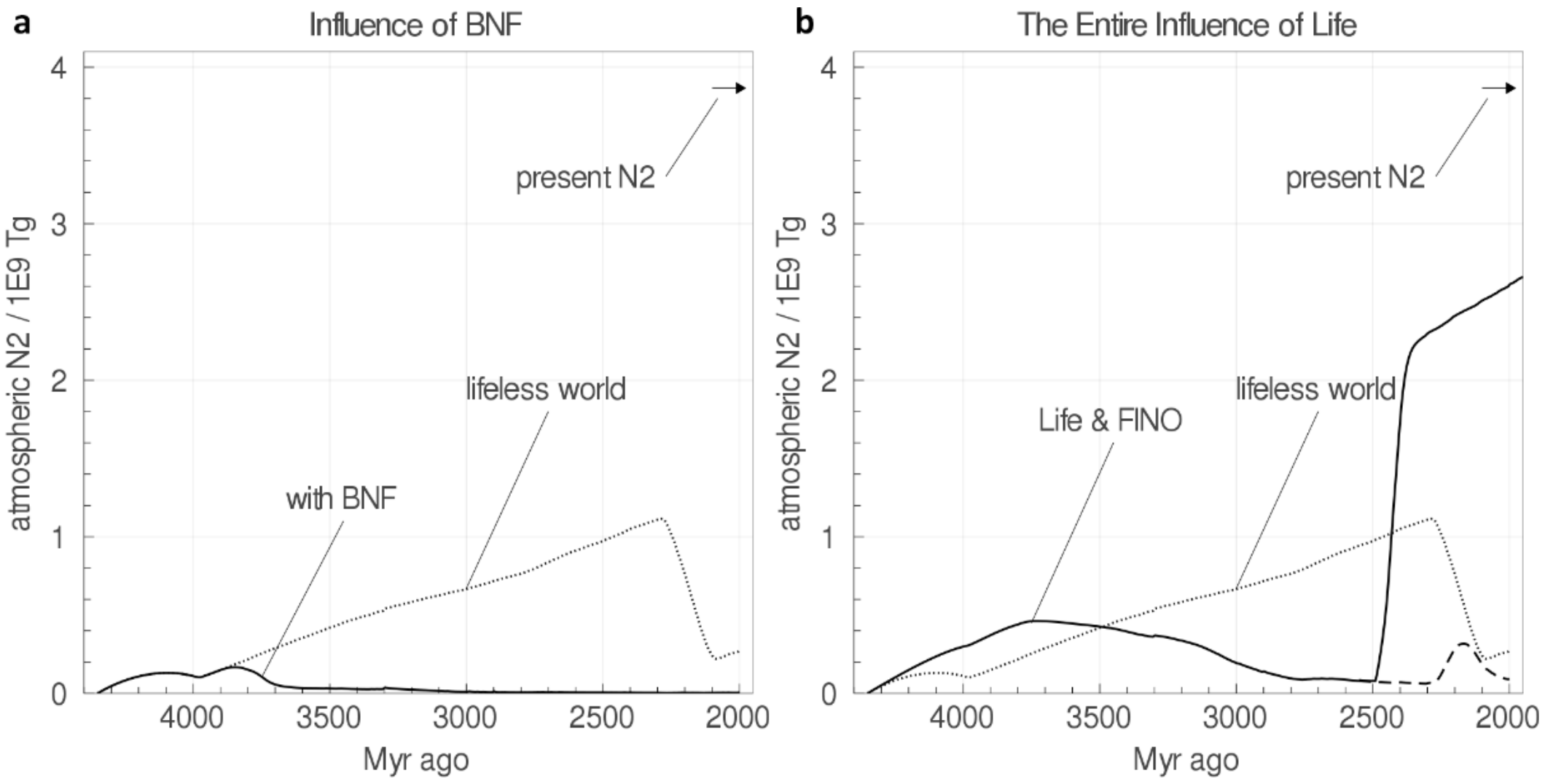

Figure 8: The impact of biological nitrogen fixation (BNF; left panel) and that of BNF in combination with $Fe^{2+}$-induced nitrogen outgassing (FINO) from the ocean and denitrification (right panel).

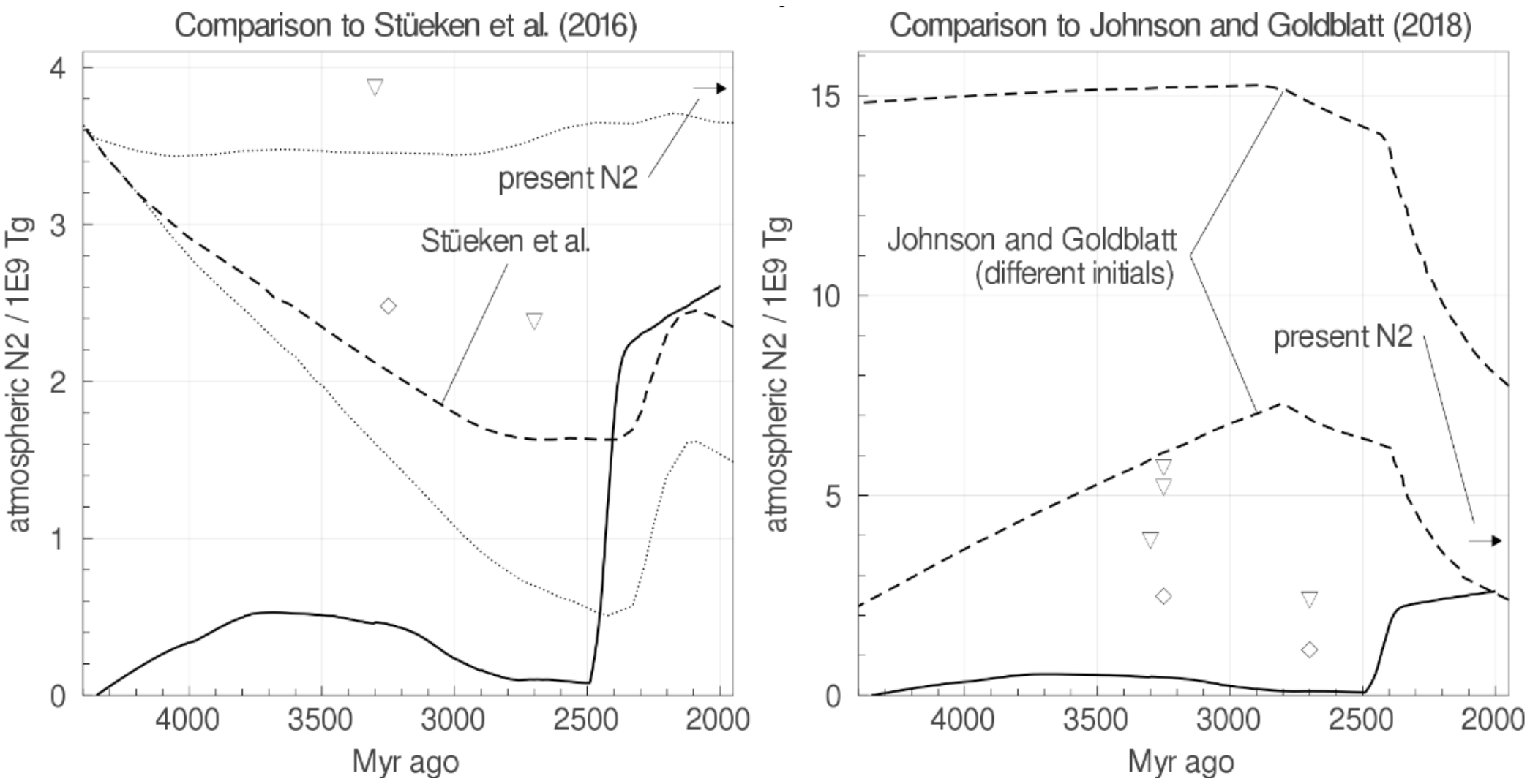

Figure 9: Comparison of the NDEV runs with zero initial nitrogen at 4350 Myr ago to other models. The solid lines are NDEV model runs as shown in Figure 8b. Diamonds ◇ and triangles ▽ indicate expectation values and upper values from geological studies for atmospheric $N_2$ pressures (see Figure 6). a: Comparison to the model run named "$F_{org}$ + Heatflow model" by Stüeken et al. [109], drawn dashed. The uncertainty range is included as dotted lines. While this alternative model does most likely not satisfy the conditions for a stable atmosphere during the late Hadean regarding escape to space, it matches the ranges of geological data. Furthermore, it also shows a strong increase of $N_2$ during the GOE transition at around 2400-2300 Myr ago. b: Comparison to the model runs "nominal run" and "nominal run with low starting atmospheric mass" by Johnson and Goldblatt [111], drawn dashed. Especially the high values of around $15 \cdot 10^9$ Tg N (equal to more than 3 bar) in the late Hadean/early Archean are highly questionable due to hardly achieved stability conditions under the solar EUV radiation (see Section 1.1). These model runs also fail the geological data and do not show an increase, but rather a massive depletion of atmospheric $N_2$ when $O_2$ accumulates in the atmosphere at 2400-2300 Myr ago.

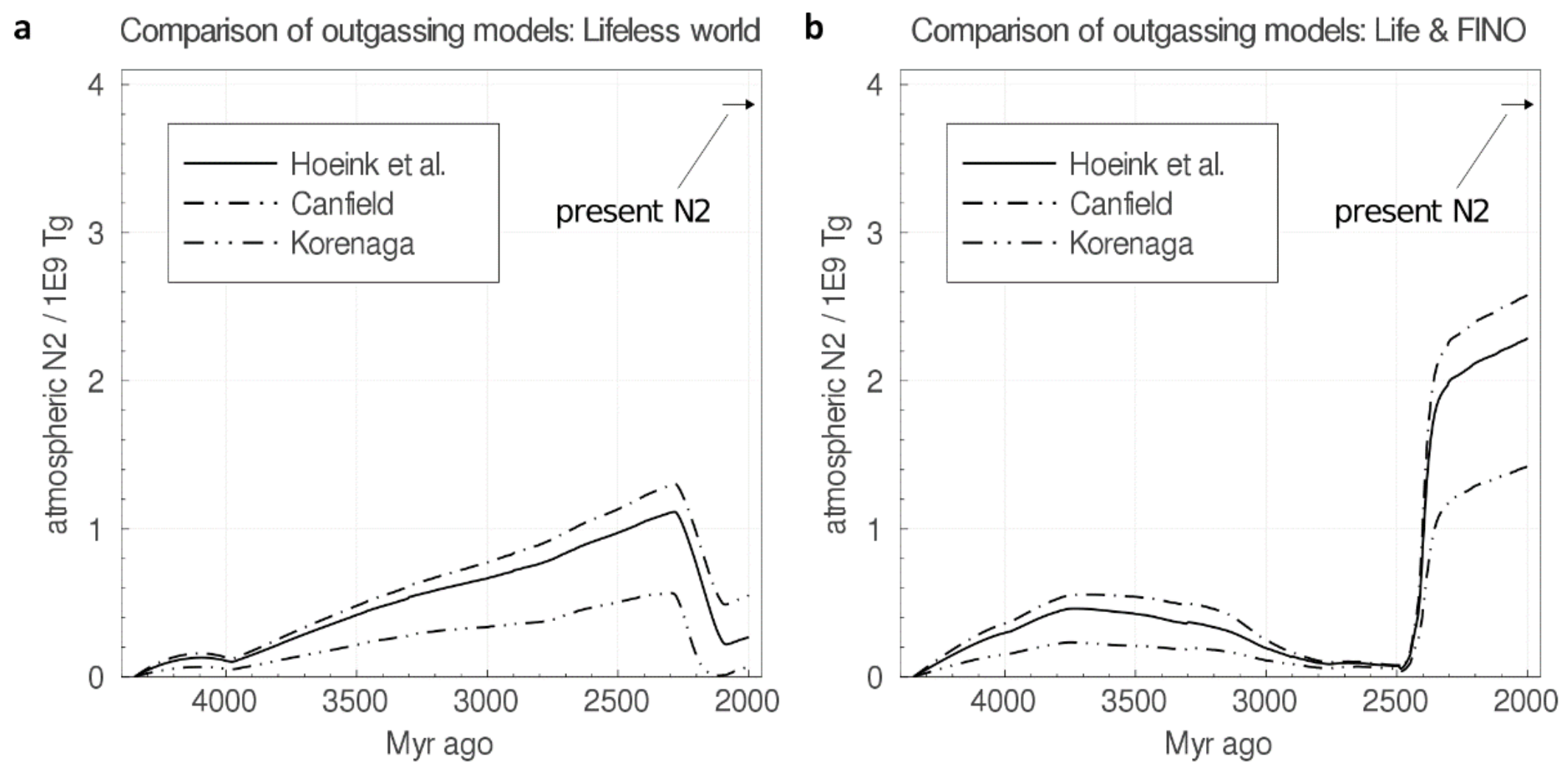

Figure 10: Comparison of different outgassing models for a lifeless world (**a**) and a scenario with FINO (**b**), as extension to the results from Figure 8. Due to the different calculation of outgassed nitrogen, the total amount of nitrogen in the upper circle varies, but the result seen from a qualitative perspective stays the same. The underlying models are taken from Canfield [112] for rather strong outgassing, Korenaga [113] for rather low outgassing and Höink et al. [114] as middle course – the last one is used as standard model for all other runs in this paper. See also Appendix Section 7.2.

## 3. $N_2$ AND $O_2$ AS GEO-BIOSIGNATURE

The detection of distinct atmospheric compounds and compositions can constitute an evidence for life. Established biosignatures are for example $CH_4$ [e.g.

29], $CH_3Cl$ [e.g. 115], $C_2H_6$ [e.g. 116], $CS_2$ [e.g. 117], $N_2O$ [e.g. 29], or $NH_3$ [e.g. 118]. Widely accepted is the requirement of larger amounts of $O_2$ especially for complex lifeforms [see 119 for a deep discussion]. The detection of $O_3$ is traceable to the presence of larger amounts of $O_2$, making $O_3$ a target for the search of life [e.g. 120, 121]. Both, liquid water and $O_2$ are certainly a requirement for life as we know it, but their origin is not (necessarily) biogenic, making it an evidence for an appropriate habitat, but not for life itself.

As derived in Section 1.3, the present Earth's atmosphere, consistent of $N_2$ and $O_2$, would be rapidly depleted when life extinguished [e.g. 54, 55]; microbes capable of denitrification contribute to retain the Earth's atmosphere. The special feature of an oxidized atmosphere with $O_2$ as bulk gas refers to long-term stable tectonic activity [122, 123] and, at least for the time after the GOE, to the influence of oxygenic photosynthesis [124, e.g. 125]. Long-term active tectonics also allows $CO_2$ depletion on a large scale [e.g. 126, 127]. These dependencies, which were discussed in the previous sections, let an $N_2$-$O_2$-atmosphere be a geo-biosignature (geo: plate tectonics, bio: aerobic lifeforms). The related theoretical background was discussed in detail by Lammer et al. [69].

But why is it so interesting to find aerobic lifeforms? At some point in the Earth's history, lifeforms turned from simple unicellular microbes to multicellular nucleus-containing organisms. It is not clear if the limiting factor was the origin of complex multicellularity or, more commonly assumed, that of developing a nucleus enclosed within membranes, which is the step from prokaryotic to eukaryotic cells. The earliest undisputed evidences for the existence of eukaryotes, found as fossils, are dated to around 1600 Myr ago [e.g. 128–130]. Eukaryotes need oxygen to live, which means they require a habitat with sufficient available oxygen beside nitrogen as nutrients. All large animal and plant cells on Earth are eukaryotic, which means that finding $N_2$ and $O_2$ indicates the existence of at least the precursors of intelligent life.

## 4. DISCUSSION AND OBSERVATIONS

In order to raise the probability of finding an actual Earth-like habitat, one needs to find terrestrial planets orbiting a G-type star rather than M- or K-type star (see discussion in Section 1.3). The most promising upcoming instrument capable of finding such bodies is PLATO, which is scheduled to be in service by late 2020s. Further, space coronagraphs such as HABEX [131] and LUVOIR [132], will be designed to directly image planets in habitable zones of G-stars and measure their respective emission spectra. To get an overview of future missions that focus on finding of habitable exoplanets, the review by Fujii et al. [133] is recommended.

### 4.1 Biosignatures in an $N_2$-$O_2$-dominated atmosphere

The uncovering of $N_2$-atmospheres remains challenging. Due to the lack of a transitional dipole moment, $N_2$ has no absorption features in the NIR and VIS spectra, and either its dissociation cross section in the EUV nor the collisionally-induced absorption line in the FIR are likely to be remotely detectable [134].

So far, only a few molecules consisting of N, namely $N_2O$ and NO, are considered as possible biosignatures and none of them are directly linked to complex lifeforms, but their presence implies the sufficient supply of N and O, which gives hints to the atmospheric bulk gas composition. The frequently discussed detection of $N_2O$ is complicated since its infrared absorption bands are very weak [135, 136, e.g. 137] and the transition bands of $N_2O$ and $NO_2$ in the visible and near-UV spectrum are overlain by strong $O_2$ and $O_3$ lines [138]. Furthermore, especially in a humid atmosphere, water vapor, $CO_2$, and even $CH_4$ overexpose $N_2O$ bands [139].

However, there is a possibility to find $N_2$, even the presence of $N_2$ and $O_2$ at once, by dimers. In transmission spectra, $N_2 \cdot O_2$-dimer absorption can be detected at 1.26 μm [140]. Further, Misra et al. [141] showed that $O_2 \cdot O_2$ dimer absorption can, together with $O_2$ vibration-rotation bands, provide information about the atmospheric pressure of exoplanets and coincidently proves the presence of $O_2$. Whether a similar analysis would be also possible with the $N_2 \cdot O_2$ dimer remains unknown so far. Both, the $O_2 \cdot O_2$ and the $N_2 \cdot O_2$ dimer, are located at 1.29 μm. Unfortunately, the $N_2 \cdot N_2$ dimer located at 4.3 μm is overlapping with a strong $CO_2$ absorption line. Schwieterman et al. [134] brought up the idea of using this dimer not only for detecting nitrogen itself, but for estimating $CO_2$ abundance, turning the disadvantage of the overlapping information into a plus. Such a detection can be successful especially in the case of high $N_2$ surface pressures of 0.5 bar and more [134]. Another indication could be Rayleigh scattering which is a constant feature of transparent $N_2$-$O_2$ atmospheres and gives the Earth the typical blue-planet appearance [142], but there is no related detection method in sight to this day.

NO can also be considered as secondary biosignature [143], more precisely as secondary metabolic biosignature [139]. For instance, it can be detected by the MIRI spectrograph (James Webb Space Telescope) at 5.3 μm [121]. It is important to point out that radical NO is a direct indicator of an $N_2$-$O_2$-dominated atmosphere because its formation is a consequence of the presence of molecular nitrogen and oxygen as the main atmospheric constituents in the planetary atmosphere. From current studies of the Earth's upper and middle atmosphere it is known that radical NO plays an important role in the structure and energetics of the upper atmosphere. NO is a heteronuclear molecule and therefore emits efficiently in the infrared, which is an important source of radiative cooling in the upper atmosphere. If transported to lower altitudes, NO will catalytically destroy ozone. Its atmospheric chemistry is characterized by a set of chemical reactions involving both $N_2$ and $O_2$. For example, the odd nitrogen chemistry - $N(^4S)$, $N(^2D)$, and NO, - is described by a simple set of chemical reactions in the Earth’s upper atmosphere [144, see e.g. , 145]:

$$N(^4S) + O_2 \rightarrow NO + O \text{ (with reaction barrier of } E_b \approx 0.3 \text{ eV)} \quad (1)$$

$$N(^2D) + O_2 \rightarrow NO + O \text{ (no barrier)} \quad (2)$$

$$N(^4S, ^2D) + NO \rightarrow N_2 + O \quad (3)$$

One can see that nitric oxide is formed in chemical reactions of dissociation products of $N_2$ with $O_2$, and is lost in collisions with atomic nitrogen $N(^4S,\ ^2D)$ reproducing $N_2$ molecules. The production and loss of NO clearly depends on the relative amounts of excited and ground state atomic nitrogen. This simple chemical set is a stiff set of thermal chemical reactions and it is usually extending (Figure 11) by additional reactions involving suprathermal nitrogen atoms [144] and the products of dissociative ionization by solar soft X-rays in photochemical models of the Earth's upper atmosphere [146, 147].

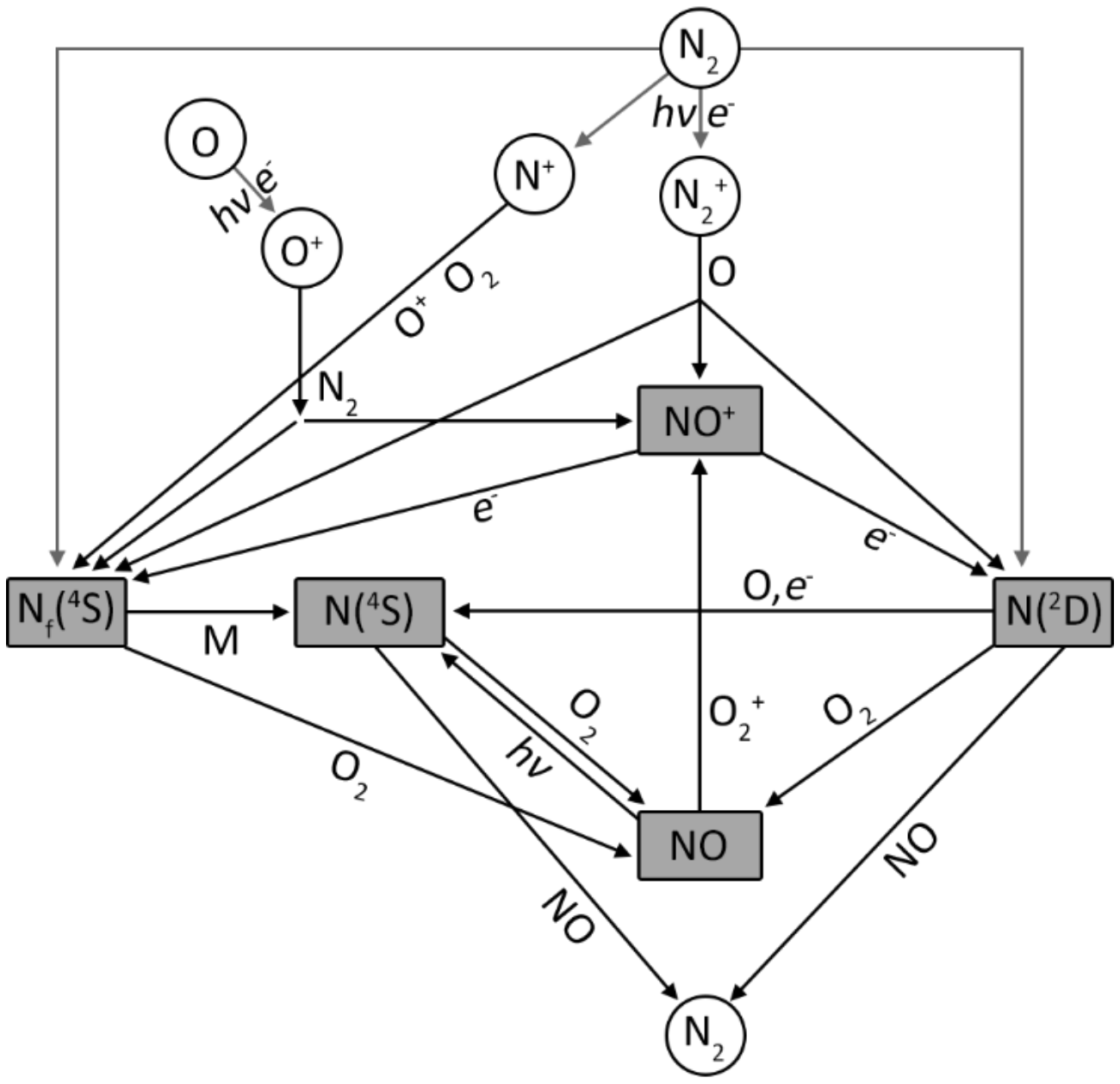

Figure 11: Schematic diagram of photochemical processes resulting in the formation of suprathermal N($^4$S) atoms and controlling nitric oxide in the Earth's upper and middle atmosphere [146, 147]. Dotted lines indicate the ionization and dissociation of $N_2$ by solar soft X-ray and ultraviolet photons and by the accompanying flux of photoelectrons and/or of auroral electrons. M denotes the elastic collisions.

Observations of NO emissions in planetary atmospheres are usually made in the ultraviolet and in the infrared by space-based instruments. The most prominent features of NO in the ultraviolet spectrum are its gamma bands in the 205.3 - 247.9 nm wavelength range [148]. Another prominent feature of NO is its fundamental band at 5.3 μm usually observed, for example, in the thermospheric infrared airglow in the Earth's upper atmosphere from 130 km up to 190 km as obtained by different space-based instruments [149–151]. All these NO observational features indicate that observations of UV emissions from NO could be a direct indication of exoplanetary $N_2$-$O_2$ atmospheres. UV observations of NO emissions observed by space observatories such as the World Space Observatory (WSO [152]) can, therefore, be even considered to be geo-biosignatures including the evidence of active tectonics [69].

A detection of $NH_3$ would prove $N_2$ in an atmosphere, but would also indicate that this atmosphere is reducing, which conflicts with $O_2$ as bulk gas. Getting its abundance is nevertheless of interest, even if it only constitutes a trace gas. Although if not representing a biosignature, the detection of $O_2$ and OH, for example by their infrared bands at 1.27 μm and 1.66 μm [121], would be helpful to prove the existence of oxygen and water in the atmosphere, which tightens the existence of an Earth-like atmosphere.

Finally, a scenario with atmospheric $N_2$ and $O_2$, but with $CO_2$ as bulk gas should be excluded as biosignature. To measure $CO_2$ abundances, its strong transition bands can be probed [e.g. 137, 153], for example by the spectrographs attached to the ELT [e.g. 154, 155].

An overview of the most promising constituents of an $N_2$-$O_2$ atmosphere with a lower amount of $CO_2$ which may be targets for missions that search for foreign life is listed in Figure 12, as well as examples for respectively matching active or future instruments.

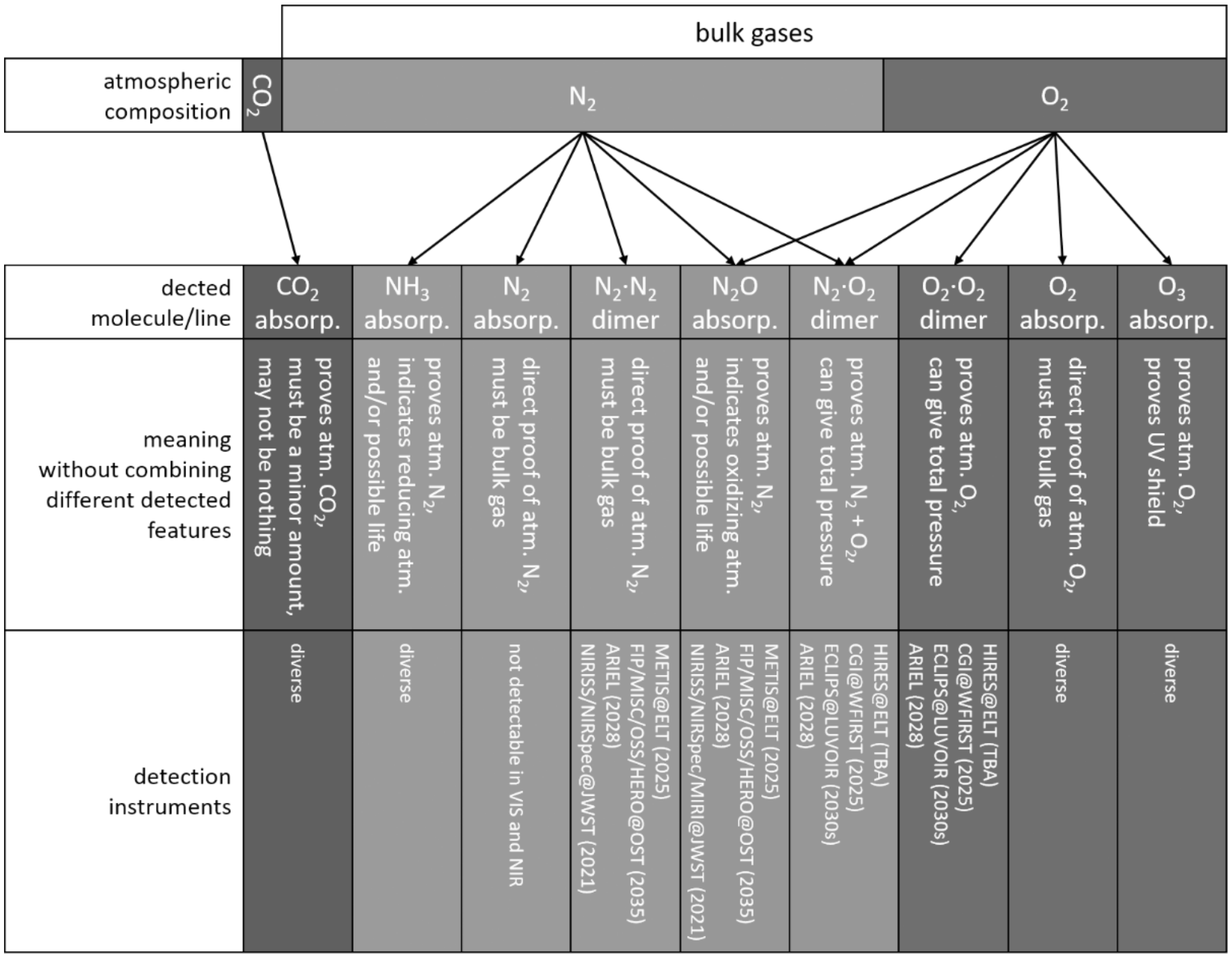

Figure 12: Overview of methods and related example instruments that may detect $N_2$-$O_2$-atmosphere indications.

## 4.2 A new focus for a habitability potential assessment

While in the past years the habitability potential of an exoplanet was estimated by its radius, mass, and distance from the respective host star, currently there is a gradual proceeding to a newer, second-estimate version for habitability assessments: the focus is now laid on the composition of its atmosphere. From recent scientific insights, we have learnt that the goodness of estimates is way higher if we have knowledge of the planet's atmospheric bulk gases, than knowing radius and mass. The potential of finding complex life is still higher when we match a suitable atmosphere and miss radius or mass, than the other way around. Furthermore, the habitable zone was recently narrowed by adding exterior conditions for $N_2$ and $CO_2$. This helps to choose the most promising candidates for probing atmospheric compositions, but, indeed, if one finds a suitable $CO_2$ amount in an $N_2$-$O_2$ dominated atmosphere on a planet outside these borderlines, one should clearly not abolish further research attempts.

## 5. CONCLUSION

Our considerations strongly point to atmospheres consisting of both $N_2$ and $O_2$ being rare due to chemical or aeronomical/astrophysical instabilities. While the latter can be circumvented in the environment of a G-type star, a chemically stable $N_2$-$O_2$-mixture can only be obtained by strong return fluxes of both species. As we could show, and as far as we know, the only complex capable of providing such massive return fluxes are aerobic lifeforms. Therefore, detecting an Earth-like atmosphere on an extrasolar planet orbiting its respective habitable zone around a G-type star is at least a strong indication for an evolved aerobic extraterrestrial biosphere. With the confirmed and proposed space and ground-based observatories, we will be able to detect such atmospheres in the near future, which will bring us a step closer to the answer on how unique the Earth's aerobic lifeforms are.

## 6. ACKNOWLEDGEMENTS

L. Sproß, M. Scherf, and H. Lammer acknowledge support by the Austrian Science Fund (FWF) NFN project S11601-N16, “Pathways to Habitability: From Disks to Active Stars, Planets and Life” and the related FWF NFN subprojects S11607-N16 “Particle/Radiative Interactions with Upper Atmospheres of Planetary Bodies under Extreme Stellar Conditions”, and S11606-N16 “Magnetospheres”.

D. Bisikalo acknowledges the support of Ministry of Science and Higher Education of the Russian Federation under the grant No. 075-15-2020-780 (N13.1902.21.0039).

V. Shematovich acknowledges the support of the Russian Science Foundation (Project No. 19-12-00370) in the preparation of subsection 4.1.

## 7. APPENDIX: MODEL DESCRIPTION

The model is set up as a simple-type box model, which allows an easy reproduction of the results. All processes are shortly described in the following. The used storages are “atmosphere”, “soil”, “ocean”, “lithosphere”, “shelf sediments”, and a storage called “subducting”. The latter is a type of temporary storage for nitrogen that is being subducted, but has not reached metamorphic conditions yet. An overview about the box model and its processes is shown in Figure 13.

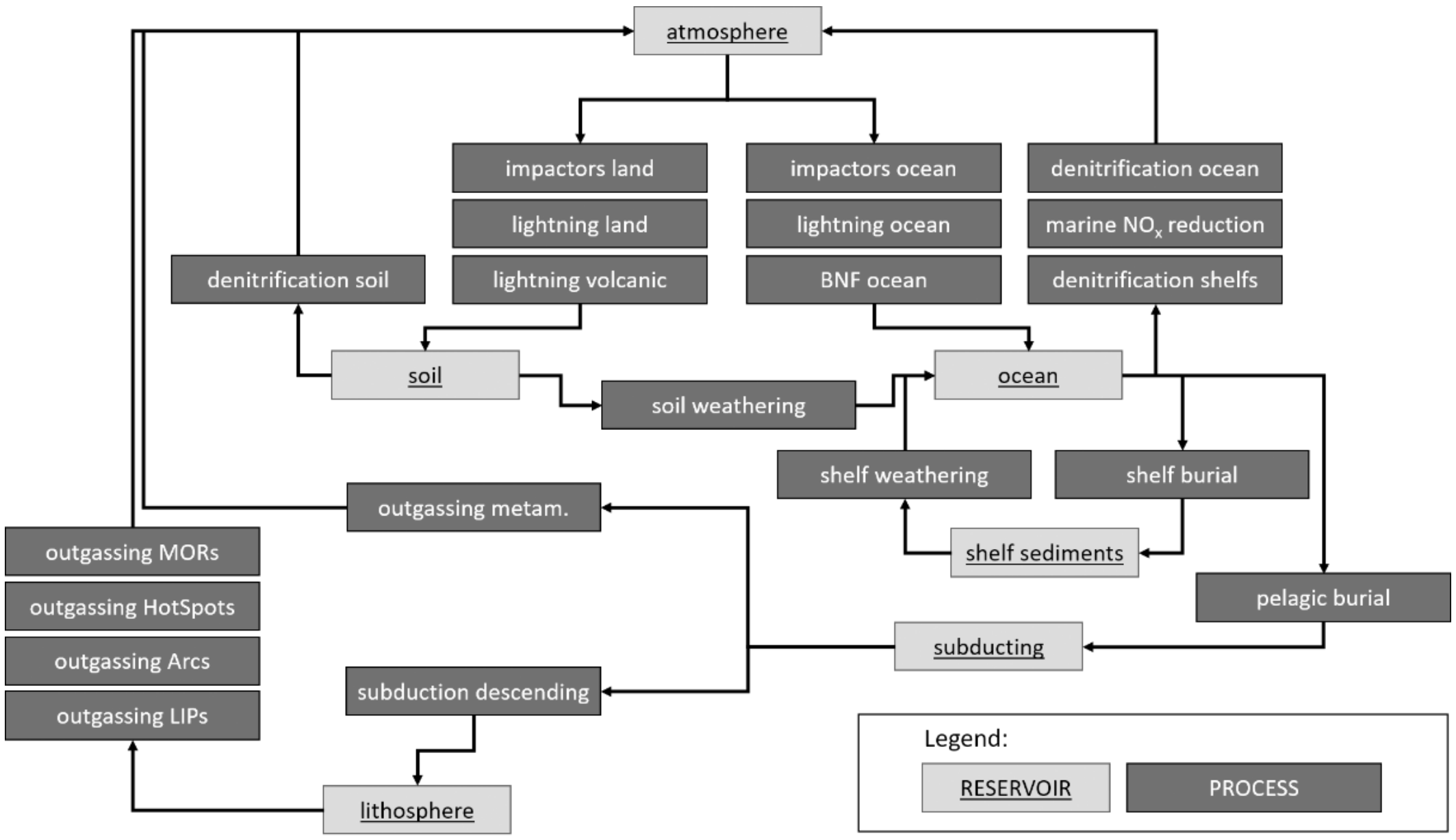

Figure 13: Structure of the used box model.

## 7.1 Initial reservoir charges

Following the argumentation in Section 2.1, the initial atmospheric $N_2$ is set to zero in the standard scenario; it is varied in Figure 7 to show the influence of its alteration. The other storages are set to zero, except for “lithosphere”, which contains $15 \cdot 10^9$ Tg initial N. The lithosphere load is chosen to approximately meet the present value under the proposed outgassing conditions.

## 7.2 Outgassing submodel

For an estimate of the outgassing potential, we used the thermal mantle evolution multiple mode model by Höink et al. [114], who considered different styles of tectonic settings.

$$Q_{\mathrm{rel}}(t) = \frac{\chi \cdot Q_{\mathrm{clas}}(T) + (1-\chi) \cdot Q_{\mathrm{slug}}(T)}{Q_{\mathrm{today}}}$$

$$Q_{\mathrm{clas}}(T) = Q_0 \left(\frac{T}{T_0}\right)^{\beta'}$$

$$Q_{\mathrm{slug}}(T) = Q_0 \left(m\frac{T}{T_0} + c\right)$$

Here, $Q_{\mathrm{rel}}$ is the heat flow relative to today, whereby $t$ is the time (ago) in Gyr, $Q_{\mathrm{today}}$ is the heat flow of today, $Q_0$ is a heat flow reference. The parameters $\beta'$ = 12, $m$ = -1.49, and $c$ = 2.18 are derived from rheology and from mantle fluid dynamics [β: 156, m&c: 157]. Following Höink et al. [114], $Q_0$ is set to 37 TW, $T_0$ (present average mantle temperature) is set to 1623 K while the initial temperature is set to 1773 K, and the influence ratio between the two tectonic regimes $\chi$ is set to 0.35. The Urey ratio, which is Ur = H/Q is assumed to be 0.35 [see 113 for a discussion]. To scale the outgassing rate at spreading zones, a formula by Turcotte and Schubert [158] is used, who argued for a direct proportionality of the outgassing and plate velocity; plate sizes are not considered here [see 113, Sec. 4.2]:

$$Q \propto T_i\sqrt{\mathrm{v}_{\mathrm{plate}}} \Rightarrow A_{\mathrm{MORs}} \propto \mathrm{v}_{\mathrm{plate}} \propto \left(\frac{Q}{T_i}\right)^2.$$

The present arc volcanism release of $N_2$ is set to $\mathbf{R}_{\mathbf{O,Arcs}}$ = 1013 Tg N/yr, while that of spreading zones is set to $\mathbf{R}_{\mathbf{O,MORs}}$ = 0.1064 Tg N/yr. The percentage that originates from previous subduction and metamorphism at arc sites is presently 42.7% [159]. It is assumed that the continental reworking rate has a large influence on the metamorphism, which is regarded by using the continental volume percentage $\xi(t)$ and the reworking rate $\varrho_{\mathrm{rw}}$ evaluated by Dhuime et al. [160]. The resulting rate can then be calculated as follows:

$$\mathbf{R}_{\mathbf{O,Arcs,mantle}}(t) = \mathbf{R}_{\mathbf{O,Arcs}} \cdot (1 - 0.427) \cdot \varrho_{\mathrm{rw}}(t) \cdot \xi_{\mathrm{oceanic}}(t) \cdot \frac{M_{\mathrm{mantle}}(t)}{M_{\mathrm{mantle}}}.$$

Hereby, $M_{\mathrm{mantle}}$(t) is the mass of stored nitrogen within the mantle at a given time respectively today. $\xi_{\mathrm{oceanic}}(t)$ describes the oceanic crust percentage, derived from the continental percentage $\xi(t)$. Reworking is discussed in Section 7.6.

Hotspot nitrogen outgassing is set to $\mathbf{R}_{\mathbf{O,Hotspots}}$ = 0.5 Tg N/yr [159, 161]. These values follow the study by Lammer et al. [69, Table A4]. The volcanic activity $A_{\mathrm{HSs}}$ is implemented after a study by Zhong [162] as follows:

$$A_{\mathrm{HSs}} \propto N \cdot v_{vert} \cdot \underbrace{V_{plume}}_{\propto R_{plume}^2} \Leftrightarrow A_{\mathrm{HSs}} \propto N \cdot \mathrm{Ra}^{\beta - n}.$$

With the heat flux $Q$ from the spreading zone model, the Rayleigh number Ra can be calculated:

$$\mathrm{Ra} = \mathrm{Nu}^{1/\beta} = \left(\frac{QD}{kA(T_i - T_s)}\right)^{1/\beta}$$

$$\Rightarrow \mathrm{Ra} = d \cdot \left(\frac{Q}{T_i - T_s}\right)^{1/\beta}$$

The variable $\beta$ is set to 0.3 [163, e.g. 164], the present Rayleigh number is set to Ra = $10^8$, and 15 plumes are assumed to be actively outgassing today. With a constant $c$ that scales the activity to the present value, the outgassing activity can be evaluated by using

$$A_{\mathrm{HSs}} = c \cdot N \cdot \mathrm{Ra}^{\beta - n}.$$

Large igneous provinces (LIPs) have to be considered for two reasons: Outgassing and enhanced volcanic lightning. In the present study, a dataset by Ernst et al. [165] is taken as basis. The outgassing rate is calculated from the nitrogen density in basalts of 800 kg N/km$^3$ via

$$\mathbf{R}_{\mathbf{LIPs}}(t)\left[\frac{\mathrm{Tg\ N}}{\mathrm{yr}}\right] = 800\ \left[\frac{\mathrm{kg\ N}}{\mathrm{km}^3}\right] \cdot 10^{-9} \cdot \frac{V_{\mathrm{LIP}}(t)}{\Delta t}\ \left[\frac{\mathrm{km}^3}{\mathrm{yr}}\right].$$

Here, $V_{\mathrm{LIP}}$ is the LIP volume and $\Delta t$ the time span in which it erupts. Missing volume information is complemented by using the surface area and a thickness of 10 km, or – when not possible – by assuming 2860000 km$^3$; these numbers are based on the averages of the existing data. Missing durations are set to 7 Myr, which is the average of the superplume eras from a study by Abbott and Isley [166].

For comparison and for revealing the uncertainty that lies within a chosen outgassing model, three approaches for spreading zone outgassing were tested:

- a polynomial fit as a rather high outgassing scenario proposed by Canfield [112], which was used for example by Claire et al. [167] or Gebauer et al. [97] with proportionality $A_{MORs} \propto Q$;
- as already described above, a Monte Carlo model developed by Höink et al. [114], which allows different convection modes during the Earth's evolution, which serves as standard scenario for the plots in Section 2.2, with proportionality $A_{MORs} \propto Q^2/T^2$ (parameters used here: $\beta' = 9.0$, $T_{start} = 1500$ °C, $Ur_0 = 0.35$, $\chi = 0.35$, see [114], their scenario in fig 3b).
- a low-amount outgassing scenario based on a model by Korenaga [113] with proportionality $A_{MORs} \propto Q$ (parameters used here: $Ur_0 = 0.30$, no continental growth, see [113] figures 47 and 51).

The above outgassing scenarios are plotted against each other in Figure 14 in terms of their outgassing potential relative to today. For a detailed discussion of different models and outgassing potential proportionalities, please see [102, Section 5.2]. . Mention that these approaches differ in the plate spreading region outgassing, arc and hotspot sites are calculated independently.

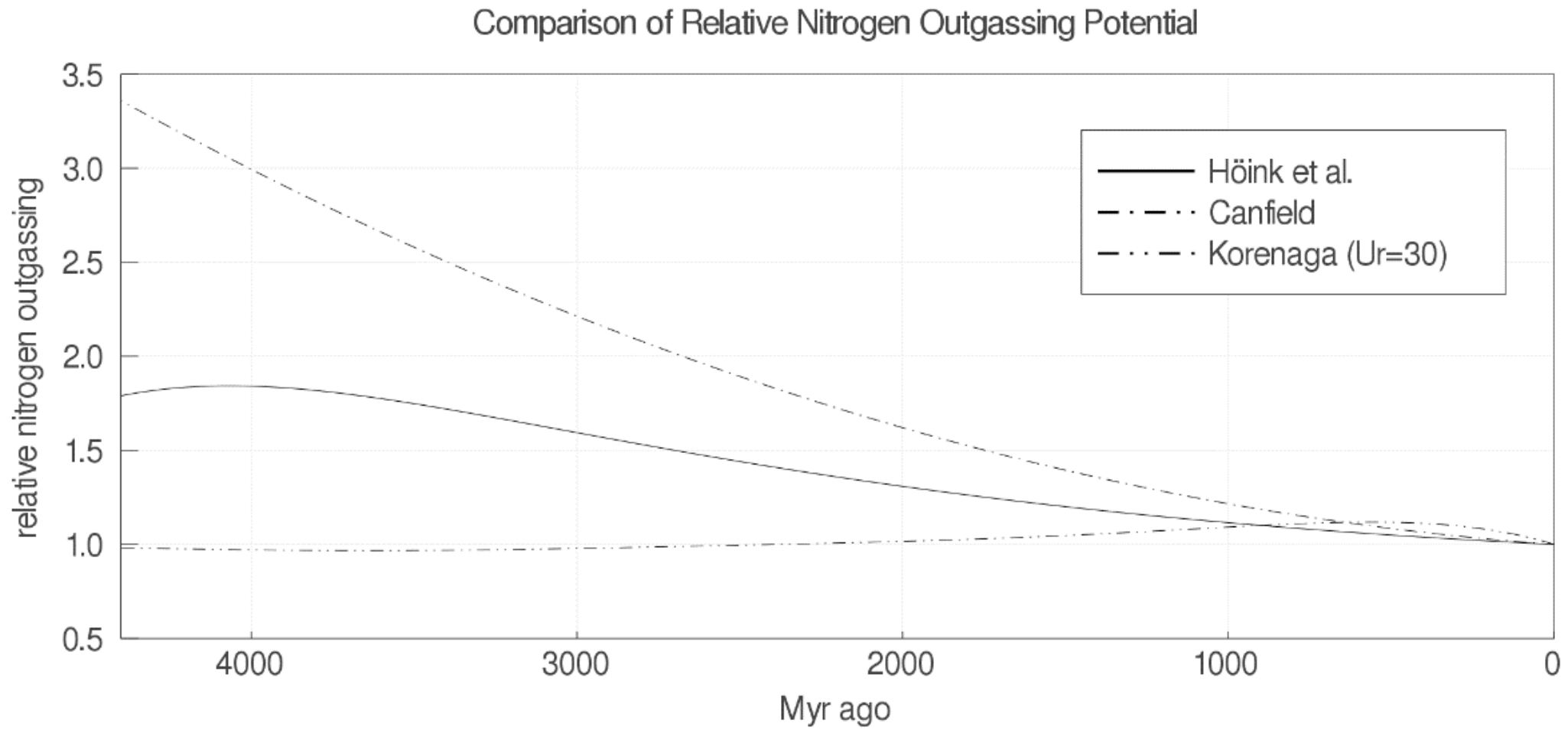

Figure 14: Comparison of the here studied outgassing potentials (see text).

## 7.3 Lightning

For the present global lightning fixation, the rate of 5 Tg N/yr [69, appendix] is divided into a marine ($\mathbf{R}_{L,o}$) and a continental ($\mathbf{R}_{L,c}$) part; the ratio between the flash rate is set to 1:10 [168], which leads to the following rates:

$$\mathbf{R}_{\mathbf{L,o}}(t = 0) = 0.45 \text{ Tg N/yr}$$

$$\mathbf{R}_{\mathbf{L,c}}(t = 0) = 4.55 \text{ Tg N/yr}$$

The alteration of the continental area follows the study by Dhuime et al. [160] who modeled continental crust volumes based on a Hf model age analysis.

Lightning fixation is dependent on atmospheric composition. While in oxidized atmospheres, $N_2$ and $O_2$ are transformed into NO, in reduced atmospheres either NO or HCN can be formed, depending on the $CO_2$/$CH_4$ amounts. In the present model, the fixation potential is dependent on $O_2$ percentage $\chi_{O2}$ and $CO_2$ percentage $\chi_{CO2}$. To do so, there is no chemical model implemented, but fits of previous studies. For early Earth, $O_2$ might be a side not, but when approaching the GOE while having a low total atmospheric pressure, its influence can grow and has to be considered. For $\chi_{O2}$, Yung et al. [169] found a more or less constant NO yield for $\chi_{O2} < 10^{-3}$; for larger values, the NO production can be approximated by

$$\mathbf{R}_{\mathbf{NO/O_2}}\left[\frac{\mathrm{Tg}}{\mathrm{yr}}\right] = \left(79 - 54 \cdot e^{-\frac{\chi O2}{0.46}} - 24 \cdot e^{-\frac{\chi O2}{0.03}}\right) \cdot 10^{11}.$$

This formula gives about 4.5 Tg N/yr for present $\chi_{O2}$. For $\chi_{O2} < 10^{-3}$, the atmosphere is considered as reducing and the fixation potential follows a study by Navarro-Gonzalez [170], who investigated the behavior of nitrogen relative to $\chi_{CO2}$. For $\chi_{O2} \leq 0.06$, the HCN production starts to contribute after Navarro-Gonzalez [170] and reaches full potential at $\chi_{O2} = 0.01$. The NO production in reduced conditions, maintained by oxygen supply from $CO_2$, works until a $CO_2$ mixing ratio of less than $\chi_{CO2} = 10^{-3}$ is reached. At very high $CO_2$ levels, NO production decreases due to a lack of $N_2$; the same problem arises when $O_2$ or a combination of $CO_2$ and $O_2$ are the predominant species in an atmosphere. To address this issue, the NO production decrease at high $CO_2$ levels from Navarro-Gonzalez [170] was used and it was found that it can be easily approximated by

$$\eta_{\mathrm{fix}} = 0 + 7.69 \cdot \chi_{\mathrm{x}} - 15.08 \cdot \chi_{\mathrm{x}}^2 \;\; \forall\, \chi_{\mathrm{x}} < 0.25.$$

Here, $\chi_x$ is the atmospheric mixing ratio of x = $N_2$. For $\chi_{N2} = 0$, the efficiency $\eta_{fix}$ is zero and $\chi_{N2} = 0.25$ yields $\eta_{fix} = 1$. Since this decrease of fixation potential is part of the NO production relative to $CO_2$, it is only necessary to modify the NO production in the case of large $O_2$ amounts. It is now important to note that most uncertainties here are not relevant for the clearly reducing conditions during the Archean, but rather for the time of the GOE transition. In the here shown approach, $CH_4$ is assumed to be less than 1% of atmospheric pressure and can therefore be neglected. This was not necessarily the case on early Earth and might be considered in the future.

Finally, one can calculate the lightning fixation rates for ocean and continents from the fixation potential:

$$\mathbf{R}_{\mathbf{L,c}}(t) = \mathbf{I}_{\mathbf{L,c}} \cdot A \cdot \xi(t) \cdot 0.293 \cdot \Phi_{\mathrm{L}},$$

$$\mathbf{R}_{\mathbf{L,o}}(t) = \mathbf{I}_{\mathbf{L,o}} \cdot A \cdot (1 - \xi(t) \cdot 0.293) \cdot \Phi_{\mathrm{L}}.$$

Hereby, $A$ is the planetary surface area, $\xi$(t) the continental crust percentage, and $\mathbf{I}_{L,c}$ and $\mathbf{I}_{L,o}$ are the fixation fluxes calculated as follows:

$$\mathbf{I}_{\mathbf{L,c}} = \mathbf{R}_{\mathbf{L.c}} \frac{1}{A \cdot 0.293} = 30.45 \cdot 10^{-3} \ \frac{\mathrm{g\,N}}{\mathrm{yr\,m^2}},$$

$$\mathbf{I}_{\mathbf{L,o}} = \mathbf{R}_{\mathbf{L,o}} \frac{1}{A \cdot 0.707} = 1.25 \cdot 10^{-3} \ \frac{\mathrm{g\,N}}{\mathrm{yr\,m^2}}.$$

The present fixation related to volcanic discharges $\mathbf{R}_{\mathbf{L,v}}$ is assumed to be 0.02 Tg N/yr. The volcanic lightning fixation in the Hadean/Archean is calculated by comparing the present value to the outgassing potential from Section 7.2.

Moreover, the impact of LIPs is regarded in relation to the ocean/land cover, because undersea eruptions do (mostly) not contribute to lightning fixation. The fixation rate is calculated as follows:

$$\mathbf{R}_{\mathbf{L,v}}(t) = \left(\frac{\mathbf{R}_{\mathbf{O,Arcs,mantle}}(t) + \mathbf{R}_{\mathbf{O,Hotspots}}(t)}{\mathbf{R}_{\mathbf{O,Arcs}} + \mathbf{R}_{\mathbf{O,Hotspots}}} \cdot \mathbf{R}_{\mathbf{L,v}}\right) + \left(\frac{\mathbf{R}_{\mathbf{O,LIPs}}(t)}{\mathbf{R}_{\mathbf{O,Arcs}}} \cdot \xi(t) \cdot 0.293 \cdot \mathbf{R}_{\mathbf{L,v}}\right).$$

## 7 . 4 Impactors

Fixation through impactor shock heating was estimated to be 0.9 Tg N/yr at 3.8 Gyr ago [70]. It is assumed that impactor fixation efficiency behaves similar to that of lightning. Here, the flux investigated by Morbidelli et al. [171] is used.

$$\mathbf{R}_{\mathbf{I}}(t) = \frac{F_{I,rel}(t)}{Z}, \qquad Z = \frac{0.9 \text{ Tg N/yr}}{F_{I,rel}(t = 3.8 \text{ Gyr ago})}$$

In this calculation, *Z* serves as normalization factor as discussed previously. The rates for land ($\mathbf{R}_{\mathbf{I,c}}$) and ocean ($\mathbf{R}_{\mathbf{I,o}}$) are then

$$\mathbf{R}_{\mathbf{I,c}}(t) = \mathbf{R}_{\mathbf{I}}(t) \cdot A \cdot \xi(t) \cdot 0.293,$$

$$\mathbf{R}_{\mathbf{I,o}}(t) = \mathbf{R}_{\mathbf{I}}(t) \cdot A \cdot (1 - \xi(t) \cdot 0.293).$$

Hereby, *A* is the planetary surface area and $\xi$(t) the continental crust percentage evaluated by Dhuime et al. [160].

Interplanetary dust and comets delivered about 0.01 Tg N/yr to the early Earth [71]. Previous studies assumed the time period in which such a delivery was significant to end at 3.8 Gyr ago [172, 173]. Recently, Morbidelli et al. [171] showed that only before 4.0 Gyr ago, a larger number of comets was available to contribute to the nitrogen budget. The fixation potential is about 2 orders higher, making delivery only a side effect. This delivery brings only fixed nitrogen to the surface, which is rapidly incorporated in sediments, so this process can be regarded by simply increasing marine or sedimentary initial contents. The amount of outgassed, fixed, and sedimented nitrogen is way higher, thus delivery by comets is here assumed as not very important.

### 7.5 Sedimentation

For sedimentation of $NH_4^+$, a Freundlich approach is used [174, 175]:

$$\Gamma = K_F(T) \cdot C^{1/n(T)}$$

$$C = \frac{M_{\text{ocean}}[\text{mol}]}{V_{\text{ocean}}}$$

Hereby, $\Gamma$ describes the sedimented material percentage, $K_F(T)$ the adsorption capacity relative to the temperature set to $K_F = 10^{-3}$ $m^3$/kg [175], $C$ the dissolved concentration in mM, $n(T)$ a constant parameter set to $n = 1$ [174, 175], $M_{\text{ocean}}$ the marine nitrogen mass, and $V_{\text{ocean}}$ the ocean volume set to constant $1.335 \cdot 10^{18}$ $m^3$ [176].

The percentage of pelagic sedimentation is 3.8% [109, 177], which must have decreased over time due to continental crust growth. Only pelagic sediments can be subducted later. Assuming $20.6 \cdot 10^6$ Tg of present marine nitrogen [178], the sedimented nitrogen mass is calculated with $M_N$ as molar mass of N via

$$m_{N,sed}[\mathrm{kg}] = \frac{\Gamma\left[\frac{\mathrm{mol}}{\mathrm{kg}}\right] \cdot \mathrm{M_N}\left[\frac{\mathrm{kg}}{\mathrm{mol}}\right]}{1.027 \cdot 10^{-9}\left[\frac{1}{\mathrm{kg}}\right]}.$$

$$\mathbf{R_{B,pelagic}}(t) = m_{N,sed}(t) \cdot 0.038 \cdot \phi(t)$$

$$\mathbf{R_{B,shelfs}}(t) = m_{N,sed}(t) \cdot (1 - 0.038) \cdot \phi(t)$$

$$\phi(t) = \frac{1 - [\xi(t) \cdot (1 - A_{\mathrm{ocean}})]}{A_{ocean}}$$

Here, $\mathbf{R_{B,pelagic}}$ is the pelagic burial rate, $\mathbf{R_{B,shelfs}}$ the shelf burial rate, $\xi(t)$ the continental crust volume again, and $A_{ocean}$ the planet ocean cover percentage.

## 7.6 Subduction and reworking

Oceanic crust and pelagic sediments are assumed to be recycled within 100 Myr [179, 180]. Following a uniform distribution, the sedimentary reworking timescale τ is then 200 Myr. Starting from an average trench-arc-distance of 166 km [181] and an average plate velocity of 5 cm/yr, it lasts about 3.3 Myr for pelagic sediments to enter the zone where metamorphism appears. The ratio of recycled to further subducted nitrogen is about 19:6 [159]. In the end, the recycled ($\mathbf{R_{O,rec,Arcs}}$) and further descending nitrogen ($\mathbf{R_{descend}}$) rates are calculated after

$$\mathbf{R_{O,rec,Arcs}}(t - 3.32\mathrm{Myr}) = 0.76 \cdot \sum_{t'=t-\tau}^{t} \frac{\mathbf{R_{B,pelagic}}(t')}{\tau} \cdot \mathrm{yr},$$

$$\mathbf{R_{descend}}(t - 3.32\mathrm{Myr}) = 0.24 \cdot \sum_{t'=t-\tau}^{t} \frac{\mathbf{R_{B,pelagic}}(t')}{\tau} \cdot \mathrm{yr}\,.$$

This simple submodel does not regard a different type of early subduction dynamics, which might be improved in the future.

## 7 . 7 Biological nitrogen fixation and denitrification

Here, the stromatolite record by Stal [182] is used, which shows the stromatolite amount $\phi_{\mathrm{Stm.}}$ relative to present day. In respect to the present day BNF rate of 61 Tg N/yr [69], the BNF rate over time $\mathbf{R_{BNF,o}}$ can be calculated by

$$\mathbf{R_{BNF,o}}(t) = \mathbf{R_{BNF,o,today}} \cdot \phi_{\mathrm{Stm.}}(t)$$

To allow model runs with extremely low $N_2$ partial pressures, the BNF rate is capped by the following restriction for N contents smaller than $m_{\mathrm{crit}}$ = 80 mbar:

$$\mathbf{R_{BNF,o}}(t) = \begin{cases} \mathbf{R_{BNF,o}} \cdot \phi_{\mathrm{Stm.}}(t) & \forall\, m_{\mathrm{N2,atm}}(t) > m_{\mathrm{crit}} \\ \mathbf{R_{BNF,o}} \cdot \phi_{\mathrm{Stm.}}(t) \cdot \dfrac{m_{\mathrm{N2,atm}}(t)}{m_{crit}} & \forall\, m_{\mathrm{N2,atm}}(t) < m_{\mathrm{crit}} \end{cases}.$$

Using this BNF-implementation, one should keep in mind that small nitrogen amounts during the time of active BNF might be not correct or unrealistic, because the atmospheric nitrogen might be completely gone, even if the model result does show a small nitrogen percentage.

### 7.8 Nitrogen oxide reduction through $Fe^{2+}$

The marine nitrogen oxide reduction is implemented in a simple way, since this process is still under debate. Here, $\sigma$ is a parameter that determines the influenced nitrogen amount.

$$\mathbf{R}_{\mathbf{RedOut}}(t) = \frac{m_{\mathrm{N2,ocean}}}{\sigma}, \qquad \sigma = 10^5$$

In the future, hopefully, more constraints might be available to implement this process more correctly.

### 7.9 $CO_2$ and $O_2$ profiles

In Figure 7, $CO_2$ and $O_2$ profiles are shown. Following the data by different studies [96, 101, 104, 183], three different $CO_2$ profiles are used and compared here (Figure 7a). One following the lower boundaries of the studies, one mean case, and one upper boundary case.

The $O_2$ profile is taken from the study by Lyons et al. [110].

## 8. AUTHOR DISCLOSURE STATEMENT

No competing financial interests exist.